\documentclass[10pt]{article}
\usepackage{amsmath}
\usepackage{latexsym}
\usepackage{amssymb}
\usepackage{amsfonts}
\usepackage{amsthm}
\title{Cauchy, Cosserat, Clausius,\\ Maxwell, Weyl Equations Revisited}
\author{J.-F. Pommaret \\ CERMICS, Ecole des Ponts ParisTech, France \\
 jean-francois.pommaret@wanadoo.fr \\
 ORCID: 0000-0003-0907-2601}
\date{  }
\begin{document}
\maketitle

\noindent
{\bf ABSTRACT}  \\

The Cauchy stress equations (1823), the Cosserat couple-stress equations (1909), the Clausius virial equation (1870), the Maxwell/Weyl equations (1873,1918) are among the most famous partial differential  equations that can be found today in any textbook dealing {\it separately and/or successively} with elasticity theory, continuum mechanics, thermodynamics, electromagnetism and electrodynamics. Over a manifold of dimension $n$,  their respective numbers are $n, n(n-1)/2, 1, n$ with a total of $(n+1)(n+2)/2$, that is $15$ when $n= 4$ for space-time. As a matter of fact, this is just the number of parameters of the Lie group of conformal transformations with $n$ translations, $n(n-1)/2$ rotations, $1$ dilatation and $n$ highly non-linear elations introduced by E. Cartan in $1922$. The purpose of this short but difficult paper is to prove that the form of these equations only depends on the structure of the conformal group for $n\geq 1$ arbitrary because they are described {\it as a whole} by the (formal) adjoint of the first Spencer operator existing in the Spencer differential sequence. Such a group theoretical implication is obtained for the first time by totally new differential geometric methods. Meanwhile, these equations can  be all parametrized by the adjoint of the second Spencer operator through $ n(n^2 - 1)(n+2)/4$ potentials.This result brings the need  to revisit the mathematical foundations of Electromagnetism and Gauge Theory according to a clever but rarely quoted paper of H. Poincar\'{e} (1901).  \\

\vspace{3cm}

\noindent
{\bf KEY WORDS}  \\

\noindent
Janet sequence; Spencer sequence; Poincar\'{e} sequence; Gauge sequence; \\
Lie group of transformations; Lie pseudogroup of transformations; \\
Conformal group of transformations; Dilatation; Elations; \\
Adjoint representation.  \\

\newpage

\noindent
{\bf 1) INTRODUCTION}  \\

Let $G$ be a Lie group with coordinates $(a^{\rho})=(a^1, ..., a^p)$ acting on a manifold $X$ with a local action map $y=f(x,a)$. According to the {\it second fundamental theorem} of Lie, if ${\theta}_1,...,{\theta}_p$ are the infinitesimal generators of the effective action of a lie group $G$ on $X$, then $[{\theta}_{\rho},{\theta}_{\sigma}]=c^{\tau}_{\rho\sigma}{\theta}_{\tau}$ where the $c=(c^{\tau}_{\rho\sigma}= - c^{\tau}_{\sigma\rho})$ are the {\it structure constants} of a Lie algebra of vector fields which can be identified with ${\cal{G}}=T_e(G)$ the tangent space to $\cal{G}$ at the identity $e\in G$ by using the action. Equivalently, introducing the non-degenerate inverse matrix $\alpha={\omega}^{-1}$ of right invariant vector fields on $G$, we obtain from crossed-derivatives the {\it compatibility conditions} (CC) for the previous system of partial differential (PD) equations called {\it Maurer-Cartan equations} or simply MC equations, namely: \\
\[ \fbox{ $ \frac{ \partial {\omega}^{\tau}_s}{\partial a^r} - \frac{\partial {\omega}^{\tau}_r}{\partial a_s} + c^{\tau}_{\rho \sigma} {\omega}^{\rho}_r {\omega}^{\sigma}_s  = 0  $  } \]
({\it care to the sign used}) or equivalently $[{\alpha}_{\rho},{\alpha}_{\sigma}]=c^{\tau}_{\rho\sigma} {\alpha}_{\tau} $ (See [9, 13, 21, 23] for more details).  \\
 
 Using again crossed-derivatives, we obtain the corresponding {\it integrability conditions} (IC) on the structure constants and the Cauchy-Kowaleski 
 theorem finally provides {\it the third fundamental theorem of Lie} saying that, for any Lie algebra $\cal{G}$ defined by structure constants $c=(c^{\tau}_{\rho\sigma})$ satisfying :\\
\[ \fbox{ $  c^{\tau}_{\rho\sigma}+c^{\tau}_{\sigma \rho}=0, \hspace{1cm} c^{\lambda}_{\mu\rho}c^{\mu}_{\sigma\tau}+c^{\lambda}_{\mu\sigma}c^{\mu}_{\tau\rho}+c^{\lambda}_{\mu\tau}c^{\mu}_{\rho\sigma}=0 $ } \]
one can construct an analytic group $G$ such that ${\cal{G}}=T_e(G)$ by recovering the MC forms from the MC equations.\\

\noindent
{\bf EXAMPLE  1.1}: Considering the affine group of transformations of the real line $y=a^2x+a^1$, the orbits are defined by $x=a^2x_0+a^1$, a definition leading to $dx=da^2x_0+da^1$ and thus $dx=((1/a^2)da^2)x+(da^1-(a^1/a^2)da^2)$. We obtain therefore ${\theta}_1={\partial}_x, {\theta}_2= x {\partial}_x \Rightarrow [{\theta}_1,{\theta}_2]= {\theta}_1$ and 
${\omega}^1= da^1-(a^1/{a^2})da^2, {\omega}^2 = (1/a^2) d a^2 \Rightarrow d{\omega}^1  + {\omega}^1\wedge{\omega}^2 = 0, d {\omega}^2 = 0  \Leftrightarrow 
[{\alpha}_1,{\alpha}_2]={\alpha}_1$ with $ {\alpha}^1 = {\partial}_1, {\alpha}_2= a^1{\partial}_1+a^2{\partial}_2$ in the following diagram:
\[ \alpha \circ \omega =  
\left( \begin{array}{cc}
1 & 0 \\
a^1 & a^2 
\end{array} \right) \circ
\left( \begin{array}{cc}
1 & 0  \\
- \frac{a^1}{a^2} &  \frac{1}{a^2}
\end{array} \right)  = 
\left( \begin{array}{cc} 
1 & 0 \\
0 & 1 
\end{array} \right)    \]

Now, if $x=a(t)x_0+b(t)$ with $a(t)$ a time depending orthogonal matrix ({\it rotation}) and $b(t)$ a time depending vector ({\it translation}) describes the movement of a rigid body in ${\mathbb{R}}^3$, then the projection of the {\it absolute speed} $v=\dot{a}(t)x_0+\dot{b}(t)$ in an orthogonal frame fixed in the body is the so-called {\it relative speed} 
$a^{-1}v=(a^{-1}\dot{a}) x_0 + (a^{-1}\dot{b})$ and the kinetic energy/Lagrangian is a quadratic function of the $1$-forms $A=(a^{-1}\dot{a}$, $a^{-1}\dot{b})$. Meanwhile, taking into account the preceding example, the {\it Eulerian speed}  $v=v(x,t)=(\dot{a}a^{-1}) x+(\dot{b}-\dot{a}a^{-1}b)$ is a quadratic function of the 1-forms $B=(\dot{a}a^{-1}, \dot{b}-\dot{a}a^{-1}b)$. We notice that $a^{-1}\dot{a}$ and $\dot{a}a^{-1}$ are both $3\times 3$ skew-symmetric time depending matrices that may be quite different.\\

\noindent
{\bf REMARK  1.2}: An easy computation in local coordinates for the case of the movement of a rigid body shows that the action of the $3\times 3$ skew-symmetric matrix $\dot{a}a^{-1}$ on the position $x$ at time $t$ just amounts to the vector product by the {\it vortex vector} $\vec{\omega}=\frac{1}{2}\vec{\nabla} \wedge \vec{v}$.   \\

The above particular cases, well known by anybody studying the analytical mechanics of rigid bodies, can be generalized as follows. \\
If $X$ is a manifold and $G$ is a lie group ({\it not acting necessarily on} $X$), let us consider maps $a:X\rightarrow G: (x)\rightarrow (a(x))$ or equivalently sections of the trivial (principal) bundle $X\times G$ over $X$. If $x+dx$ is a point of $X$ close to $x$, then $T(a)$ will provide a point $a+da=a+\frac{\partial a}{\partial x}dx$ close to $a$ on $G$. We may bring $a$ back to $e$ on $G$ by acting on $a$ with $a^{-1}$, {\it  either on the left or on the right}, getting therefore a $1$-form $a^{-1}da=A$ or $(da)a^{-1}=B$ with value in $\cal{G}$. As $aa^{-1}=e$ we also get $(da)a^{-1}=-ada^{-1}=-b^{-1}db$ if we set $b=a^{-1}$ as a way to link $A$ with $B$. When there is an action $y=ax$, we have $x=a^{-1}y=by$ and thus $dy=dax=(da)a^{-1}y$, a result leading through the first fundamental theorem of Lie to the equivalent formulas:\\
\[   a^{-1}da=A=({A}^{\tau}_i(x)dx^i=-{\omega}^{\tau}_{\sigma}(b(x)){\partial}_ib^{\sigma}(x)dx^i)  \]
\[   (da)a^{-1}=B=({B}^{\tau}_i(x)dx^i={\omega}^{\tau}_{\sigma}(a(x)){\partial}_ia^{\sigma}(x)dx^i)  \]
Introducing the induced bracket $[A,A](\xi,\eta)=[A(\xi),A(\eta)]\in {\cal{G}}, \forall \xi,\eta\in T$, we may define the {\it curvature} $2$-form $dA-[A,A]=F\in {\wedge}^2T^*\otimes {\cal{G}}$ by the local formula ({\it care again to the sign}):\\
\[     {\partial}_iA^{\tau}_j(x)-{\partial}_jA^{\tau}_i(x)-c^{\tau}_{\rho\sigma}A^{\rho}_i(x)A^{\sigma}_j(x)=F^{\tau}_{ij}(x)  \]
This definition can also be adapted to $B$ by using $dB+[B,B]$ and we obtain from the second fundamental theorem of Lie:\\

\noindent
{\bf THEOREM 1.3}: There is a {\it nonlinear gauge sequence}:\\
\[ \fbox{  $   \begin{array}{ccccc}
X\times G & \longrightarrow & T^*\otimes {\cal{G}} &\stackrel{MC}{ \longrightarrow} & {\wedge}^2T^*\otimes {\cal{G}}  \\
a                & \longrightarrow  &    a^{-1}da=A         &    \longrightarrow & dA-[A,A]=F
\end{array}  $ }    \eqno{(1)}   \]

In 1956, at the birth of GT, the above notations were coming from the EM potential $A$ and EM field $dA=F$ of relativistic Maxwell theory. Accordingly, $G=U(1)$ (unit circle in the complex plane) $\longrightarrow dim ({\cal{G}})=1$) was the {\it only possibility} to get a $1$-form $A$ and a $2$-form $F$ with vanishing structure constants $c=0$.  \\

Choosing now $a$ "close" to $e$, that is $a(x)=e+t\lambda(x)+...$ and linearizing as usual, we obtain the linear operator $d:{\wedge}^0T^*\otimes {\cal{G}}\rightarrow {\wedge}^1T^*\otimes {\cal{G}}:({\lambda}^{\tau}(x))\rightarrow ({\partial}_i{\lambda}^{\tau}(x))$ leading to (See again [13] for more details):\\

\noindent
{\bf COROLLARY  1.4}: There is a {\it linear gauge sequence}:\\ 
\[  \fbox{ $ {\wedge}^0T^*\otimes {\cal{G}}\stackrel{d}{\longrightarrow} {\wedge}^1T^*\otimes {\cal{G}} \stackrel{d}{\longrightarrow} {\wedge}^2T^*\otimes{\cal{G}} \stackrel{d}{\longrightarrow} ... \stackrel{d}{\longrightarrow} {\wedge}^nT^*\otimes {\cal{G}}\longrightarrow  0  $  }     \eqno{(2)}  \]
which is the tensor product by $\cal{G}$ of the Poincar\'{e} sequence for the exterior derivative. \\

 In order to introduce the previous results into a variational framework, we may consider a Lagrangian on $T^*\otimes \cal{G}$, that is an {\it action} $W=\int w(A)dx$ where $dx=dx^1\wedge ...\wedge dx^n$ and to vary it. With $A=a^{-1}da= - (db)b^{-1}$ we may introduce $\lambda=a^{-1}\delta a= - (\delta b)b^{-1}\in {\cal{G}}={\wedge}^0T^*\otimes {\cal{G}}$ with local coordinates ${\lambda}^{\tau}(x)=-{\omega}^{\tau}_{\sigma}(b(x))\delta b^{\sigma}(x)$ and we obtain in local coordinates ([13], p 180-185):
 \[ \fbox{ $   \delta A=d\lambda - [A,\lambda]  \Leftrightarrow  \delta A^{\tau}_i={\partial}_i\lambda^{\tau}-c^{\tau}_{\rho\sigma}A^{\rho}_i{\lambda}^{\sigma} $  }  \eqno{(3)}  \] 
 Then, setting $\partial w/\partial A={\cal{A}}=({\cal{A}}^i_{\tau})\in {\wedge}^{n-1}T^*\otimes {\cal{G}}^* $, we get:\\
\[  \delta W=\int {\cal{A}}\delta Adx=\int {\cal{A}}(d\lambda-[A,\lambda])dx  \]
and therefore, after integration by part, the Euler-Lagrange (EL) equations of Poincar\'{e} ([8, 13]):\\
\[   \fbox{ $   {\partial}_i{\cal{A}}^i_{\tau}+c^{\sigma}_{\rho\tau}A^{\rho}_i{\cal{A}}^i_{\sigma}=0   $ }  \eqno{(4)} \]
Such a linear operator for $\cal{A}$ has non constant coefficients linearly depending on $A$ and is the adjoint of the previous operator (up to sign). \\
However, setting now $(\delta a)a^{-1}=\mu\in {\cal{G}}$, we get $\lambda=a^{-1}((\delta a)a^{-1})a=Ad(a)\mu$ while, setting $a'=ab$, we get the {\it gauge transformation} for any $b\in G$ (See [13], Proposition $14$, p $182$):
\[ \fbox{   $  A \rightarrow A'=(ab)^{-1}d(ab)=b^{-1}a^{-1}((da)b+adb)=Ad(b)A+b^{-1}db,   \Rightarrow F' = Ad(b) F$  }  \eqno{(5)}   \] 
Setting $b=e+t\lambda+...$ with $t\ll 1$, then $\delta A$ becomes an infinitesimal gauge transformation. Finally, $ a'=ba\Rightarrow A'=a^{-1}b^{-1}((db)a+a(db))=a^{-1}(b^{-1}db)a+A\Rightarrow \delta A=Ad(a)d\mu$ when $b=e+t\mu +...$ with $t\ll 1$.
Therefore, introducing $\cal{B}$ such that ${\cal{B}}\mu= {\cal{A}}\lambda$, we get the divergence-like equations:
\[  \fbox{  $  {\partial}_i{\cal{B}}^i_{\sigma}=0  $  }  \eqno{(6)}   \]

We provide some more details on the  {\it Adjoint representation} $Ad: G \rightarrow aut({\cal{G}}): a \rightarrow Ad(a)$. which is defined by a linear map $ M(a) = ({M}^{\tau}_{\rho}(a)) \in aut({\cal{G}})$ and we have the involutive system ([13], Proposition 10, p 180):  
\[  \fbox{  $      \frac{\partial {M}^{\tau}_{\mu}}{\partial a^r} + c ^{\tau}_{\rho \sigma} {\omega}^{\rho}_r (a) {M}^{\sigma}_{\mu} = 0   $   }  \eqno{(7)} \] 
by using the fact that {\it any right invariant vector field on $G$ commutes with any left invariant vector field on $G$} (See [1, 11, 35] for applications of ({\it reciprocal distributions}) to Differential Galois Theory). In addition, as $a^{-1} \delta a = \lambda$, we obtain therefore successively:
\[    Ad(a^{-1})\lambda = a \lambda a^{-1} = a (a^{-1} \delta a ) a^{-1} = \delta a a^{-1} = \mu \, \,  \Rightarrow  \, \, d Ad(a^{-1}) \lambda = (d \delta a) a^{-1}- \delta a a^{-1} d a a^{-1}  \]
\[  \begin{array}{lcl}
 Ad(a) d Ad(a^{-1}) \lambda   &  =  &  a^{-1} (d \delta a a^{-1} - \delta a a^{-1} d a a^{-1})a   \\
        & =  &   a^{-1} d \delta a - (a^{-1} \delta a a^{-1}) d a   \\
        & =  &   a^{-1} \delta d a + (\delta a^{-1}) d a \\
        & =  &  \delta A  
        \end{array}   \]
As a byproduct, the operator $\nabla: \lambda \rightarrow d \lambda - [A, \lambda]$ only depends on $A$ and is the "{\it twist }" of the derivative $d$ under the action of $Ad(a)$ on 
${\cal{G}}$ whenever $A= a^{-1} d a$ in the gauge sequence. Setting:
\[   \nabla (\alpha \otimes \lambda) = d \alpha \otimes \lambda + (-1)^r \alpha \wedge \nabla \lambda, \forall \alpha \in {\wedge}^r T^* ; \lambda \in {\cal{G}} \]
an easy computation shows that $(\nabla \circ \nabla \lambda)^{\tau}_{ij} = c^{\tau}_{\rho \sigma} F^{\rho}_{ij} {\lambda}^{\sigma} = 0 $ and we obtain:  \\

\noindent
{\bf COROLLARY 1.5}: The following ${\nabla}$-sequence:
\[   \fbox{  $     {\wedge}^0 T^* \otimes {\cal{G}} \stackrel{\nabla} \longrightarrow {\wedge}^1T^* \otimes {\cal{G}} \stackrel{\nabla}{\longrightarrow} ... \stackrel{\nabla}{\longrightarrow} 
{\wedge}^n T^* \otimes {\cal{G}} \longrightarrow 0  $  }  \eqno(8)  \]
is {\it anothe}r locally exact linearization of the non-linear gauge sequence which is isomorphic to copies of the Poincar\'{e} sequence and describes infinitesimal gauge transformations.  \\
We have the commutative diagram:  \\
\[ \fbox{  $    \begin{array}{lcccc} 
   &   & 0 &  & 0     \\
   &   & \downarrow &  &  \downarrow \\
       \mu &  \in  & {\wedge}^0 T^* \otimes {\cal{G}} & \underset 1{ \stackrel{grad}{\longrightarrow}} & T^* \otimes  {\cal{G}} \\     
     \downarrow & Ad(a) &  \downarrow     &    &  \downarrow         \\     
     \lambda & \in & {\wedge}^0 T^* \otimes {\cal{G}} & \underset 1{ \stackrel{\nabla}{\longrightarrow }}  & T^* \otimes {\cal{G}}    \\ 
           &   & \downarrow &  &  \downarrow  \\
                  &   & 0 &  & 0                         
\end{array}  $  }   \]
both with the corresponding adjoint diagram that does not seem to be known: 
\[  \fbox{  $         \begin{array}{ccccccc}
&  & 0  &  &  0  &  &  \\
 & &   \uparrow  &  &  \uparrow   & & \\
 0  & \longleftarrow & {\wedge}^n T^* \otimes  {\cal{G}}^* & \underset 1{\stackrel{div}{\longleftarrow}} &  {\wedge}^{n-1} T^* \otimes {\cal{G}}^* & \ni & {\cal{B}}    \\  
       &     &     \uparrow   &  & \uparrow  & Ad(a)  & \uparrow     \\
  0  &  \longleftarrow  &  {\wedge}^n T^* \otimes {\cal{G}}^* &  \underset 1{\stackrel{ad(\nabla)}{\longleftarrow}} & {\wedge}^{n-1} T^* \otimes {\cal{G}}^*  &  \ni  &  {\cal{A}}  \\
 &   & \uparrow  &  &  \uparrow  &  &  \\
   &  & 0  &  &  0 
\end{array}  $  }   \]   \\   \\

{\it In a completely different local setting}, if $G$ acts on $X$ and $Y$ is a copy of $X$ with an action graph $X \times G \rightarrow X \times Y: (x, a) \rightarrow (x, y = a x = f(x,a))$, we may use the theorems of S. Lie in order to find a basis $\{ {\theta}_{\tau} \mid 1 \leq \tau \leq p=dim(G) \}$ of infinitesimal generators of the action. If $\mu = ({\mu}_1, ..., {\mu}_n)$ is a multi-index of length $\mid \mu \mid = {\mu}_1 + ... + {\mu}_n$ and ${\mu}+1_{i} = ({\mu}_1,..,{\mu}_{i-1},{\mu}_i + 1, {\mu}_{i+1} ,...,{\mu}_n)$, we may introduce the system of infinitesimal Lie equations or Lie algebroid $R_q \subset J_q(T)$ with sections defined by ${\xi}^k_{\mu} (x) = {\lambda}^{\tau} (x) {\partial}_{\mu} {\theta}^k_{\tau}(x)$ for an arbitrary section ${\lambda} \in {\wedge}^0 T^* \otimes {\cal{G}}$ and the trivially involutive operator $j_q: T \rightarrow J_q(T): \theta \rightarrow ({\partial}_{\mu} \theta , 0 \leq \mid \mu \mid \leq q)$ of order $q$. We finally obtain the {\it Spencer operator} through the chain rule for derivatives [9, 19, 38, 39]:  \\
\[   \fbox{  $  (d{\xi}_{q+1})^k_{\mu, i}(x) =   {\partial}_i{\xi}^k_{\mu}(x) -{\xi}^k_{\mu +1_i}(x) = {\partial}_i{\lambda}^{\tau}(x) {\partial}_{\mu} {\theta}^k_{\tau}(x)  $   }  \eqno(9) \]

\noindent
{\bf THEOREM 1.6}: When $q$ is large enough to have an isomorphism $R_{q+1} \simeq R_q \simeq {\wedge}^0 T^* \otimes {\cal{G}}$ and the following  {\it linear Spencer sequence} in which the operators $D_r$ are induced by $d$ as above:  \\
\[ \fbox{  $  0 \longrightarrow \Theta \stackrel{j_q}{\longrightarrow} R_q \stackrel{D_1}{\longrightarrow} T^* \otimes R_q \stackrel{D_2}{\longrightarrow} {\wedge}^2 T^* \otimes R_q \stackrel{D_3}{\longrightarrow} ... \stackrel{D_n}{\longrightarrow} {\wedge}^nT^*\otimes R_q \longrightarrow  0  $  }    \eqno(10)   \]
is isomorphic to the linear gauge sequence but with a completely different meaning because $G$ is now acting on $X$ and $\Theta \subset T$ is such that $[\Theta, \Theta ] \subset \Theta$.  \\

\noindent
{\bf EXAMPLE 1.7}: ({\it Weyl group}): For an arbitrary dimension $n$, the Lie pseudogroup of conformal transformations, considered as a Lie group of transformations, has $n$ translations, $n(n-1)/2$ rotations, 1 dilatation and $n$ nonlinear elations, that is a total of $(n+1)(n+2)/2$ parameters while the Weyl subgroup of transformations has only $(n^2 +n +2)/2$ parameters. When $n=2$ and the standard Euclidean metric, we may choose the infinitesimal generators $ {\theta_1= {\partial}_1, {\theta}_2 = {\partial }_2, \theta}_3 = x^1 {\partial}_2 - x^2 {\partial}_1, {\theta}_4 = x^1 {\partial}_1 + x^2 {\partial}_2$ of the Weyl subgroup with $2 + 1 + 1 = 4$ parameters by taking out the elations. Setting ${\xi}^k_{\mu}= {\lambda}^{\tau} {\partial}_{\mu} {\theta}^k_{\tau} $ with $\lambda = ({\lambda}^{\tau}(x))\in {\cal{G}}$, we have the $4 \times 4$ full rank matrix allowing to describe the isomorphism $R_2\simeq R_1\simeq {\wedge}^0 T^* \otimes {\cal{G}}$:
\[    {\xi}^1 = {\lambda}^1 - x^2 {\lambda}^3 + x^1 {\lambda}^4, \,\, {\xi}^2 = {\lambda}^2 + x^1 {\lambda}^3 + x^2 {\lambda}^4, \,\, {\xi}^2_1 =  - {\xi}^1_2 = {\lambda}^3 , \,\, 
{\xi}^1_1 = {\xi}^2_2 = {\lambda}^4     \]
as follows:  \\
\[  \fbox{  $  \left( \begin{array}{l}
{\xi}^1  \\
{\xi}^2  \\
 {\xi}^2_1 = - {\xi}^1_2   \\
 {\xi}^1_1 = {\xi}^2_2 = A
 \end{array} \right)  = 
\left( \begin{array}{cccc}
1 & 0 & -x^2 & x^1 \\
0 & 1 & x^1 & x^2 \\
0 & 0 & 1 & 0 \\
0 & 0 & 0 & 1
\end{array} \right)
\left( \begin{array}{c}
{\lambda}^1 \\
{\lambda}^2  \\
{\lambda}^3  \\
{\lambda}^4
\end{array} \right)    $   }   \]
Now, in order to determine $ad(D_1)$, we have to integrate by parts the duality  summation:
\[  {\sigma}^1_1 ({\partial}_1 {\xi}^1 - {\xi}^1_1) +{\sigma}^1_2 ({\partial}_1 {\xi}^2 - {\xi}^2_1) + {\sigma}^2_1 ({\partial}_2 {\xi}^1 - {\xi}^1_2) + {\sigma}^2_2 ({\partial}_2 {\xi}^2 - {\xi}^2_2)
  + {\mu}^{12,r} {\partial}_r {\xi}^2_1 + {\nu}^r {\partial}_r {\xi}^1_1  \]
in which we have taken into account the Medolaghi equations $ {\xi}^2_2 - {\xi}^1_1 =0, {\xi}^1_2 + {\xi}^2_1 =0, {\xi}^k_{ij}=0$ defining the Weyl algebroid. We get the $2$ Cauchy + $1$ Cosserat + $1$ Clausius equations describing the adjoint of the first Spencer operator in which we may have ${\sigma}^{1,2} \neq {\sigma}^{2,1}$:
\[  {\partial}_r {\sigma}^{i,r}= f^i, \,\,\,  {\partial}_r {\mu}^{12,r} + { \sigma}^{1,2}  -  {\sigma}^{2,1} = m^{12}, \,\,\, {\partial}_r{\nu}^r + {\sigma}^r_r = u  \]
that we can transform into the $4$ {\it pure divergence} equations by comparing $ad(D_1)$ with $ad(d)$:
\[  {\partial}_r ( {\sigma}^{i,r} ) = f^i, \,\,\,  {\partial}_r ({\mu}^{12,r} + x^1 {\sigma}^{2,r} - x^2 { \sigma}^{1,r} ) = m^{12} + x^1 f^2 - x^2 f^1, \,\,\, 
{\partial}_r ( {\nu}^r + x^i {\sigma}^r_i) = u + x^i f_i    \]
a result not completely evident at first sight, both with the parametrization problem that we now study and solve. In the case of the Poincar\'{e} subgroup for the Euclidean metric, the Spencer sequence is isomorphic to the tensor product of the Poincar\'{e} sequence for the exterior derivative by a Lie algebra ${\cal{G}}$ of dimension $n+ n(n-1)/2=n(n+1)/2$. The Spencer operator $D_2:T^* \otimes R_1 \rightarrow {\wedge}^2 T^* \otimes R_1$ is thus parametrized by the Spencer operator $D_1$ and the operator $ad(D_1)$ is thus parametrized by the operator $ad(D_2)$ which is providing $n^2(n^2 - 1)/4$ potentials, that is $3$ when $n=2$. As the Spencer operator $D_1$ may be defined by the $6$ equations:  \\
\[  {\partial}_1 {\xi}_1= A_{1,1}, {\partial}_1 {\xi}_2 - {\xi}_{1,2} = A_{2,1}, {\partial}_2 {\xi}_1 - {\xi}_{1,2} = A_{1,2}, {\partial}_2 {\xi}_2 = A_{2,2}, {\partial}_1 {\xi}_{1,2} = B_1, 
{\partial}_2 {\xi}_{1,2} = B_2  \]
because $R_1$ is defined by the Lie equations $({\xi}_{1,1}=0, {\xi}_{1,2} + {\xi}_{2,1}=0, {\xi}_{2,2}=0)$.  \\
The three CC made by $D_2$ are thus:  \\
\[  {\partial}_2 A_{1,1} - {\partial }_1 A_{2,1} + B_1=0, \,\,\, {\partial}_2 A_{1,2} - {\partial}_1 A_{2,2} + B_2 =0, \,\,\, {\partial}_2 B_1 - {\partial}_1 B_2 = 0  \]
Multiplying these three equations respectively by $ {\phi}^1, {\phi}^2, {\phi}^3$, summing and integrating by parts, we obtain the first order parametrization:  \\
\[  \left\{ \begin{array}{lcr}
A_{1,1} & \rightarrow & - {\partial}_2 {\phi}^1  \\
A_{1,2} & \rightarrow  & - {\partial}_2 {\phi}^2   \\
A_{2,1} & \rightarrow & {\partial}_1 {\phi}^1 \\
A_{2,2} & \rightarrow & {\partial}_1 {\phi}^2 
\end{array} \right.  \hspace{1cm}
\left\{ \begin{array}{lcl}
B_1 & \rightarrow & {\phi}^1 - {\partial}_3 {\phi}^3  \\
B_2 & \rightarrow & {\phi}^2 + {\partial}_1 {\phi}^2
\end{array} \right.  \]
\[ \fbox{  $  {\sigma}^{1,1} = - {\partial}_2 {\phi}^1, \,\, {\sigma}^{1,2} = - {\partial}_2 {\phi}^2, \,\, {\sigma}^{2,1}= {\partial}_1 {\phi}^1, \,\, {\sigma}^{2,2} = {\partial}_1 {\phi}^2, \,\, 
{\mu}^{12,1} = - {\partial}_2 {\phi}^3 + {\phi}^1, \,\, {\mu}^{12,2} = {\partial}_1 {\phi}^3 + {\phi}^2  $  } \]
a result that could not have been even imagined by the Cosserat brothers.  \\

\noindent
{\bf 2) CONFORMAL GROUP OF TRANSFORMATIONS}  \\

 The idea is to notice that the brothers are {\it always} dealing with the same group of rigid motions because the lines, surfaces or media they consider are all supposed to be in the same $3$-dimensional background/surrounding space which is acted on by the group of rigid motions, namely a group with $6$ parameters ($3$ {\it translations} + $3$ {\it rotations}). In 1909 it should have been strictly impossible for the two brothers to extend their approach to bigger groups, in particular to include the only additional {\it dilatation} of the Weyl group that will provide the virial theorem and, {\it a fortiori}, the {\it elations} of the conformal group considered later on by H.Weyl [40]. In order to emphasize the reason for using Lie equations, we now provide the explicit form of the highly nonlinear $n$ finite elations and their infinitesimal counterpart in a conformal Riemannian space of dimension $n$ with a non-degenerate metric 
 $\omega$, namely:\\
\[  \fbox{  $   y=\frac{x-x^2b}{1-2(bx)+b^2x^2}  \Rightarrow  {\theta}_s= - \frac{1}{2} x^2 {\delta}^r_s{\partial}_r+{\omega}_{st}x^tx^r{\partial}_r   \Rightarrow \  
{\partial}_r{\theta}^r_s=n{\omega}_{st}x^t, \hspace{2mm}   \forall 1\leq r,s,t \leq n   $   }  \]
where the underlying metric is used for the scalar products $x^2,bx,b^2$ involved.  \\

Our purpose is to exhibit {\it directly} the Cauchy [2], Cosserat [3], Clausius [23], Maxwell [7], Weyl [40] (CCCMW) equations by computing with full details the adjoint of the first Spencer operator $D_1: {\hat{R}}_3 \rightarrow T^* \otimes {\hat{R}}_3$ for the conformal involutive finite type third order system ${\hat{R}}_3 \subset J_3(T)$ for any dimension $n\geq 1$, in particular for $n\geq 1$ along with the results obtained in [30]. In general, one has $n$ translations (t) $+$  $n(n-1)/2$ rotations (r) +  $1$ dilatation (d) + $n$ nonlinear elations (e), that is a total of  $(n+1)(n+2) /2$ parameters, thus $15$ when $n=4$. As a byproduct, the Cosserat couple-stress equations will be obtained for the Killing involutive finite type second order system ${\hat{R}}_2 \subset J_2(T)$. It must be noticed that not even a single comma must be changed when $n=3$ when our results are compared to the original formulas provided by the bothers Cosserat in $1909$. We only need recall the specific features of the standard first order Spencer operator $d: {\hat{R}}_3 \rightarrow T^* \otimes {\hat{R}}_2$ as follows by considering the multi-indices for the $n$ translations, the $n(n-1)/2$ rotations, the only dilatation and the $n$ elations separately as follows:   \\
\[   ({\xi}^k(x), {\xi}^k_i(x), {\xi}^k_{ij}(x), {\xi}^k_{ijr}(x) = 0) \rightarrow  ({\partial}_i {\xi}^k(x) - {\xi}^k_i(x), \,\,  {\partial}_i {\xi}^k_j(x) - {\xi}^k_{ij}(x) , \,\,  {\partial}_i {\xi}^r_r(x) - {\xi}^r_{ri}(x), {\partial}_r {\xi}^k_{ij}(x) )    \]
in the {\it duality summation}:  \\
\[  \fbox{  $   {\sigma}^i_k ({\partial}_i {\xi}^k(x) - {\xi}^k_i(x) )+ {\mu}^{ij}_k ({\partial}_i {\xi}^k_j(x) - {\xi}^k_{ij}(x)) + {\nu}^i ({\partial}_i {\xi}^r_r(x) - {\xi}^r_{ri}(x)) + 
 {\pi}^{ij,r}_k ({\partial}_r {\xi}^k_{ij}(x) )  $  }    \]
We have obtained a first simplification by noticing that the third order jets vanish, that is to say ${\xi}^k_{rij}=0$. Indeed, starting with the Euclidean or Minkowski metric 
$\omega$ with vanishing Christoffel symbols $\gamma=0$, the second order conformal equations can be provided in the parametric form:  \\
 \[  \fbox{  $  {\xi}^k_{ij} = {\delta}^k_i A_j(x) + {\delta}^k_j A_i(x) - {\omega}_{ij} {\omega}^{kr} A_r(x) \Leftrightarrow {\xi}^r_{ri} = n A_i(x)  $  }   \]
The desired result follows from the fact that this system is homogeneous and it is well known that ${\hat{g}}_3=0, \forall n\geq 3$ as in [23] but also in any case as in [30]. \\
A second simplification may be obtained by using the (constant) metric in order to raise or lower the indices in the implicit summations considered. In particular, we have successively:  \\
\[     {\omega}_{rj} {\xi}^r_i + {\omega}_{ir} {\xi}^r_j - \frac{2}{n} {\omega}_{ij} {\xi}^r_r =0 \,\,\,   \Leftrightarrow   \,\,\, 
       {\omega}_{rj} {\xi}^r_i + {\omega}_{ir} {\xi}^r_i = 2 A(x) {\omega}_{ij}   \]
\[ \Rightarrow  A(x) = {\xi}^1_1(x) = {\xi}^2_2 (x) = ... = {\xi}^n_n (x) = \frac{1}{n} {\xi}^r_r (x) \,\,\, \Rightarrow \,\,\,  {\xi}_{i,j} + {\xi}_{j,i}=0, \forall i\neq j  \]
Now, we recall that we have chosen the notations in such a way that the system is formally integrable if and only if we have:  \\

\noindent
{\bf LEMMA 2.1 }: The sum of the Spencer operators ${\partial}_i A(x) - A_i(x)$ and $ {\partial}_i A_j(x) - 0$ is a linear combination of the original Spencer operators ${\partial}_i {\xi}^r _r(x) - {\xi}^r_{ri}(x) $ and ${\partial}_i {\xi}^r_{rj}(x) - 0$ and conversely.  \\

\noindent                                                                                                                                                                                          
${\it Proof}$: Substituting, we obtain successively:
\[  {\partial}_i A(x) - A_i(x) = \frac{1}{n} ({\partial}_i{\xi}^r_r - {\xi}^r_{ri})   \]
\[ {\partial}_i{A}_j(x) - 0= \frac{1}{n} ( {\partial}_i({\xi}^r_{rj} - 0) = \frac{1}{n} ( {\partial}_i{\xi}^r_{rj} - {\xi}^r_{rij})  \]

\hspace*{12cm} $ \Box  $  \\

In this situation, it follows that ${\sigma}^{i,j} {\xi}_{i,j} = {\Sigma}_{i < j}({\sigma}^{i,j} - {\sigma}^{j,i}) {\xi}_{i,j} + \frac{1}{n} {\sigma}^r_r {\xi}^r_r $ and we may set 
${\mu}^{ij}_k {\xi}^k_{ij} = - {\mu}^i A_i$ where ${\mu}^i$ is a linear (tricky) function of the ${\mu}^{ij}_k$ with constant coefficients only depending on $\omega$. The new equivalent duality summation becomes:
\[   {\sigma}^{i,r} {\partial}_r {\xi}_i + {\Sigma}_{i<j} ( {\mu}^{ij,r} {\partial}_r{\xi}_{i,j}  - ({\sigma}^{i,j} - {\sigma}^{j,i} ){\xi}_{i,j} )  - {\sigma}^r_r A(x)  -  {\mu}^i A_i(x) + {\nu}^i ({\partial}_i A (x) - A_i(x)) + {\pi}^{i,r}({\partial}_r A_i(x))      \]
and is important to notice that we may have ${\sigma}^{i,j} \neq {\sigma}^{j,i}$ when $i < j$. Integrating by parts and changing the signs, we finally obtain the Poincar\'{e} Euler-Lagrange equations in the following form that allows to avoid using the structure constants of the conformal Lie algebra:  \\

\[  \fbox{ $  {\xi}_i \longrightarrow {\partial}_r {\sigma}^{i,r} = f^i    \,\,\, (Cauchy \,\, stress \,\, equations)   $  }   \]

\[  \fbox{  $  {\xi}_{i,j}, i< j \longrightarrow  {\partial}_r {\mu}^{ij,r} + {\sigma}^{ij} - {\sigma}^{ji} = m^{ij}  \,\,\, (Cosserat \,\,  couple{-}stress \, \,equations) $ }  \]

\[  \fbox{  $  A(x) \longrightarrow {\partial}_r {\nu}^r   + {\sigma}^r_r =  u             \,\,\, (Clausius \,\, equation ) $  }  \]

\[  \fbox{  $  A_i(x) \longrightarrow         {\partial}_r {\pi}^{i,r}  + {\mu}^i + {\nu}^i =  v^i    \,\,\,( Maxwell/Weyl \,\, equations )    $   } \]   \\   \\
Transforming these equations into pure divergence-like equations as we already did for the Poincar\'{e} equations by using now the isomorphism 
$ R_q \simeq {\wedge}^0T^*\otimes {\cal{G}}$ is much more difficult.  \\

First of all, using the last example of the introduction dealing with the Weyl group, we obtain when the second order jets vanish:  \\

\[  \fbox{ $   {\partial}_r {\sigma}^{i,r} = f^i    $  }   \]

\[ \fbox{  $   {\partial}_r ({\mu}^{ij,r} + x^j {\sigma}^{i,r}  - x^i {\sigma}^{j,r}  ) = m^{ij} + x^j f^i - x^i f^j,   \forall 1   \leq i < j \leq n    $  }   \]

\[  \fbox{  $  {\partial}_r ( {\nu}^r   + x^i {\sigma}^r_i  ) =  u  +  x^i f_i   $  }  \]

\[  \fbox{  $      {\partial}_r ( {\pi}^{i,r}  + x^i  {\nu}^r +  ...  ) =  v^i   +  x^i u + ...     $   } \]

\noindent
the two first equations, introduced by the Cosserat brothers in ([3], p 137, 167), are {\it exactly} the ones used in  continuum mechanics in order to study the equilibrium of forces and couples bringing the symmetry of the stress-tensor when $\mu = 0 $ and $m=0$, where the left member is the Stokes formula applied to the total surface density of momentum while the right member is the total volume density of momentum (See [12, 13] for details).  \\

Accordingly, the only problem left is to fill in the details for the Maxwell/Weyl equations but this will be the most difficult part of this paper because it will depend on the elations !.  \\

\noindent
{\bf REMARK 2.2 }: As noticed by the Cosserat brothers themselves, a major difference existing between the Cauchy and the Cosserat elasticity theories is that the {\it compatibility conditions} (CC) for the respective {\it fields} are described by a second order operator for the first but by a first order operator for the second. When $n=2$, we have indeed for the Killing operator ${\cal{D}}$ [13, 33]: \\
\[ \fbox{  $   2 \,  {\partial}_1 {\xi}^1 = {\Omega}_{11}, \, 2  {\partial}_2 {\xi}^2 = {\Omega}_{22}, \, {\partial}_1 {\xi}^2 + {\partial}_2 {\xi}^1 = {\Omega}_{12} \,\, \Rightarrow  \, \, 
    {\partial}_{22} {\Omega}_{11} + {\partial}_{11}  {\Omega}_{22} - 2 \, {\partial}_{12} {\Omega}_{12} = 0  $  }  \]
    and for the corresponding Spencer operator $D_1$ [19, 26]:
\[ \fbox{  $   {\partial}_i {\xi}^k - {\xi}^k_i = X^k_i, \, {\partial}_i {\xi}^k_j = X^k_{j,i} \,\, \Rightarrow \,\, {\partial}_i X^k_j - {\partial}_j X^k_i + X^k_{i,j} - X^k_{j,i} = 0, 
\, {\partial}_i X^k_{l,j} - {\partial}_j X^k_{l, i} = 0   $  }  \]
Though we are dealing in both cases with the Lie pseudogroup $\Gamma$ of rigid transformations:
\[ {\omega}_{kl} (f(x)) {\partial}_i f^k(x) {\partial}_l f^l(x) = {\omega}_{ij}(x) \] 
for the Euclidean metric $\omega$, we discover that {\it the differential sequence used is not the same}. This fact has never been acknowledged by the people working on Cosserat media because this result only depends on the {\it fundamental diagram I} ([9], p 183) [12, 13, 15]. Multiplying the Riemann operator by a test function $\phi$ and integrating by parts while taking into account that ${\sigma}^{ij} {\Omega}_{ij} = {\sigma}^{11} {\Omega}_{11} + {\sigma}^{22} {\Omega}_{22} + 2 \, {\sigma}^{12} {\Omega}_{12}$, we obtain the well known second order Airy parametrization [29, 34, 36]:  \\
\[ \fbox{  $  {\sigma}^{11} = {\partial}_{22} \phi, \,\,\, {\sigma}^{22} = {\partial}_{11} \phi, \,\,\, {\sigma}^{12} = {\sigma}^{21} = - {\partial}_{12} \phi        $  }   \]
The same comment can be done on the possibility to enlarge the Lie pseudogroup $\tilde{\Gamma}$ of Weyl transformations to the Lie pseudogroup $\hat{\Gamma}$ of conformal transformations in arbitrary dimension [30]. The first difficulty is that the "{ \it Vessiot structure equations }" are superseding the "{\it Cartan structure equations}", a fact never acknowledged by Cartan and all followers up to now [12, 32] and the second is that the Spencer sequence is also largely superseding the Cartan structure equations which are using exterior calculus on jet bundles of order $q$ without any possibiity to quotient them down on the base manifold $X$ as we saw.   \\

\noindent
{\bf EXAMPLE 2.3}: ({\it Projective group of the real line}): With $n=1$ and $K=\mathbb{Q}$, let us consider the Lie pseudogroup defined by the third order Schwarzian OD equation with standard jet notations:   \\
\[     \frac{y_{xxx}}{y_x} - \frac{3}{2} (\frac{y_{xx}}{y_x})^2 = 0   \,\,\,\,  \Rightarrow  \,\,\,\,   {\xi}_{xxx}=0  \]
A basis of infinitesimal generators is $\{ {\theta}_1 = {\partial}_x, \, {\theta}_2 = x \, {\partial}_x, \, {\theta}_3 = \frac{1}{2} x^2 \}$ and we have the following diagram in which the columns of the $3\times 3$ matrix describes the components of $j_2(\theta)$:  \\ 
\[ \fbox{  $   \left( \begin{array}{lcl}
{\xi} &  &   \\
{\xi}_x & =  &  A  \\
 {\xi}_{xx} &  =  & A_x
\end{array} \right)   =
\left( \begin{array}{cccccc}
1 & x & \frac{1}{2} x^2   \\
0 & 1 & x   \\
 0 & 0 & 1
\end{array} \right)
\left( \begin{array}{c}
{\lambda}^1 \\
{\lambda}^2  \\
{\lambda}^3  
\end{array} \right)   $  }  \]

\noindent
In order to construct the adjoint of the first Spencer operator $D_1: {\hat{R}}_3  \rightarrow T^* \otimes {\hat{R}}_3$ when there are only one translation, no rotation but only one dilatation and  only one elation, we have to consider the duality sum with ${\xi}_{xxx} = 0$:  \\
\[   \sigma \, ({\partial}_x \xi - {\xi}_x) + \nu \, ({\partial}_x {\xi}_x - {\xi}_{xx}) +  \pi \, ({\partial}_x {\xi}_{xx} - 0)  \]
Integrating by parts and changing the sign, we get the board of first order operators allowing to define $ad(D_1)$, namely the Cauchy/Cosserat stress equation, the Clausius virial equation and the Maxwell/Weyl equation successively:  \\
\[ \fbox{  $   \left\{  \begin{array}{lclcc}
\xi & \longrightarrow & {\partial}_x \sigma & = & f  \\
{\xi}_x & \longrightarrow & {\partial}_x \nu + \sigma & = & u  \\
{\xi}_{xx} & \longrightarrow & {\partial}_x \pi + \nu & = & v  
\end{array} \right.  $  }  \]
We may obtain therefore the pure divergence equations:  \\
\[   \fbox{  $  \left\{ \begin{array}{lcl}
{\partial}_x (\sigma) & = &  f  \\
{\partial}_x ( \nu + x \, \sigma ) & = &  x \, f  + u \\
{\partial}_x ( \pi + x \nu + \frac{1}{2} x^2 v) & = & \frac{1}{2} x^2 f + x \, u +  v
\end{array}  \right.  $  }  \]
Coming back now to the Janet sequence in the following {\it Fundamental Diagram I} of [30]:  \\
\[ \fbox{  $  \begin{array}{rccccccccc}
  & && & &0 & & 0 &  &  \\
   & & & & & \downarrow & & \downarrow & &  \\
   & 0 &\longrightarrow &\Theta &\stackrel{j_3}{\longrightarrow} & 3 & \underset 1{\stackrel{D_1}{\longrightarrow} } & 3 &\longrightarrow 0 & \hspace{5mm} Spencer   \\
   & & & & & \downarrow & & \parallel  &  &  \\
    & 0 & \longrightarrow &1 &\stackrel{j_3}{\longrightarrow} & 4 & \underset 1{\stackrel{D_1}{\longrightarrow} }& 3 & \longrightarrow 0&   \\
    & & & \parallel & & \hspace{3mm}\downarrow \Phi & & \downarrow & &  \\
    0\longrightarrow & \Theta & \longrightarrow & 1 & \underset 3{\stackrel{{\cal{D}}}{\longrightarrow} } & 1 & \longrightarrow & 0 & & \hspace{5mm} Janet  \\
    & & & & & \downarrow & & & &  \\
     & & & & & 0 & & & & 
 \end{array}  $  }  \]
we notice that the central row splits because $j_3$ is an injective operator and the corresponding sequence of differential modules is the splitting sequence $ 0 \rightarrow D^3 \rightarrow D^4 \rightarrow D \rightarrow 0 $. It follow that the adjoint sequence also splits in the following commutative and exact adjoint diagram:  \\
\[ \fbox{  $   \begin{array}{ccccccccc}
    & & & & 0 & & 0 &  &  \\
    & &  & & \uparrow & & \uparrow &  \\
    & & 0 & \longleftarrow & 3 & \underset 1{\stackrel{ ad(D_1) }{\longleftarrow} } & 3 &  &    \\
    & & \uparrow & & \uparrow & & \parallel  &  &  \\
 0 & \longleftarrow & 1 &\stackrel{ ad(j_3 ) }{\longleftarrow} & 4 & \underset 1{\stackrel{ ad(D_1) }{\longleftarrow} }& 3 & \longleftarrow  & 0   \\
    & & \parallel & & \uparrow  & & \uparrow  & &  \\
 0 & \longleftarrow & 1 & \underset 3{\stackrel{ad({\cal{D}})}{\longleftarrow} } & 1 &  &  0  & &  \\
    & &\uparrow & & \uparrow & & & &  \\
    & & 0 & & 0 & & & & 
 \end{array} $  }   \]
In a more effective but quite surprising way, the kernel of $ad(D_1)$ in the adjoint Spencer sequence is defined by the successive differential conditions:  \\
\[  (f=0, \, u=0, \, v=0 ) \Rightarrow ({\partial}_x \, \sigma = 0, \, {\partial}_x \,  \nu + \sigma =0, \, {\partial }_x \, \pi + \nu = 0 ) \Rightarrow  \fbox{ $ {\partial}_{xxx} \, \pi =0  $ }  \]
It is thus isomorphic to the kernel of $ad({\cal{D}})$ in the adjoint of the Janet sequence. But ${\cal{D}}$ is a Lie operator in the sense that ${\cal{D}}\xi= 0, {\cal{D}} \eta =0 \Rightarrow {\cal{D}} [\xi, \eta] =0$ or, equivalently, $[ \Theta, \Theta ] \subset \Theta$. It follows that the " stress " appearing in the Cauchy operator which is the adjoint of the Lie operator ${\cal{D}}$ in the Janet sequence has {\it strictly nothing to do}  with the " stress " appearing in the Cosserat couple-stress equations provided by the adjoint of $D_1$ appearing in the Spencer sequence. This confusion, which is even worst that the ones we have described at length in the many recent references (See [9,12, 25, 32] for details and letters on the controversy Cartan/Vessiot and [24, 29, 34, 36] for details on the controversy Beltrami/Einstein), leads to revisit the mathematical foundations of both continuum mechanics and general relativity. \\

It is important to point out that $ad(j_3)$ is nothing else than the Euler-Lagrange equation in higher dimensional variational calculus (See chapter $VIII$ part A of [13] and the last chapter of [16] for details on the variational sequence). More precisely, we have:  \\
\[ j_3: (\xi = A, {\partial}_x \xi = B, {\partial}_{xx} \xi = C, {\partial}_{xxx} \xi = D) \, \, \Rightarrow \,\, 
D_1:  ({\partial}_x A - B = E, {\partial}_x B - C = F, {\partial}_x C - D = G) \] 
Multiplying by three test functions $({\lambda}^1, {\lambda}^2, {\lambda}^3)$ and integrating by parts while changing the sign as usual, we obtain the adjoint operator $ad(D_1)$ with CC the adjoint operator $ad(j_3)$:   \\
\[  ({\partial}_{xxx} {\mu}^4 - {\partial}_{xx} {\mu}^3 + {\partial}_x {\mu}^2 - {\mu}^1 = \nu ) \, \, \Leftarrow  \, \, ({\partial}_x {\lambda}^1 = {\mu}^1, {\partial}_x {\lambda}^2 + {\lambda}^1 = {\mu}^2, {\partial}_x {\lambda}^3 + {\lambda}^2 = {\mu}^3, {\lambda}^3= {\mu}^4)  \] 
The central vertical short exact splitting sequence is described by:  \\
\[ {\mu}^4 \rightarrow ((0, 0, 0, {\mu}^4), \, \,  ({\mu}^1, {\mu}^2, {\mu}^3, {\mu}^4) \rightarrow ({\mu}^1, {\mu}^2, {\mu}^3) = ( \sigma, \nu, \pi)  \] 
a result that could not be even imagined without these new methods.  \\

\noindent
{\bf EXAMPLE 2.4}: ({\it Conformal Group}) When $n=2$, the conformal group has $6$ parameters and we should follow the same procedure after adding the two elations: \\
\[ {\theta}_5= \frac{1}{2} ( (x^1)^2 - (x^2)^2) {\partial}_1 + x^1 x^2 {\partial}_2, \,\, \, {\theta}_6  = x^1 x^2 {\partial}_1 + \frac{1}{2} ((x^2)^2 - (x^1)^2) {\partial}_2   \]
 in such a way that ${\partial}_r {\theta}^r_5= 2 x^1, {\partial}_r {\theta}^r_6=2x^2$ (See [17] for the relation with the Clausius virial theorem and [18] for the relation with the conformal group).  The only difference is that we have now to deal with the $6$ right members $(f^1, f^2, m^{12}, u, v^1, v^2)$. As we have been only using the Spencer bundles $C_0$ and $C_1$, these results have strictly nothing to do with $C_2 $ involving $2$-forms and the so-called Cartan curvature, a result also proving that the mathematical foundations of Gauge theory must be {\it totally} revisited as we have no longer any link with the unitary group $U(1)$ [21]. \\

With more details, when $n=2$, the Lie equations are (See [30) for more details):\\
\[  {\xi}^1_2 + {\xi}^2_1 = 0, \,\, {\xi}^2_2 - {\xi}^1_1 = 0, \,\, {\xi}^1_{22}- {\xi}^1_{11} = 0, \,\, {\xi}^2_{22} - {\xi}^1_{12} = 0, \,\, {\xi}^1_{12} - {\xi}^2_{11} = 0, \,\, 
{\xi}^2_{12} - {\xi}^1_{11} = 0, \,\, {\xi}^k_{ijr} = 0  \]
and we may now add the two previous elations in order to obtain similarly the $6$ parametric sections ${\xi}_2(x) $ of the jet bundle ${\hat{R}}_2 \subset J_2(T)$ from the arbitrary sections $\lambda = \lambda(x)$. We have the following diagram in which the columns of the $6 \times 6 $ matrix are the components of $j_2(\theta)$:\\
\[ \fbox{   $   \left( \begin{array}{l}
{\xi}^1  \\
{\xi}^2  \\
{\xi}^2_1 = - {\xi}^1_2 \\
{\xi}^1_1 = {\xi}^2_2 = A  \\
{\xi}^1_{11} = {\xi}^2_{12} = - {\xi}^1_{22} = A_1  \\
{\xi}^1_{12} = {\xi}^2_{22} = - {\xi}^2_{11} = A_2
\end{array} \right)   =
\left( \begin{array}{cccccc}
1 & 0 & -x^2 & x^1 &  {\theta}^1_5 & {\theta}^1_6    \\
0 & 1 & x^1 & x^2 & {\theta}^2_5  & {\theta}^2_6     \\
0 & 0 & 1 & 0 & x^2 &  - x^1  \\
0 & 0 & 0 & 1 & x^1 & x^2   \\
0 & 0 & 0 & 0 & 1 & 0  \\
0 & 0 & 0 & 0 & 0 & 1
\end{array} \right)
\left( \begin{array}{c}
{\lambda}^1 \\
{\lambda}^2  \\
{\lambda}^3  \\
{\lambda}^4  \\
{\lambda}^5  \\
{\lambda}^6
\end{array} \right)   $   }  \]

Using the fact that we have now ${\xi}^k_{ijr}=0$ because the system is homogeneous and we may assume that ${\hat{g}}_3=0$ as in [30], we have the new duality summation:
\[ \fbox{  $  {\sigma}^i_k ({\partial}_i {\xi}^k - {\xi}^k_i) + {\mu}^{12,r} ({\partial}_r {\xi}^2_1 - {\xi}^2_{1r}) + {\nu}^r ({\partial}_r A - A_r) + {\pi}^{i,r} {\partial}_r A_i $  } \]

Multiplying on the left by the row vector $(f^1, f^2, m^{12}, u, v^1, v^2)$, we obtain for the four columns on the left the same results that only involve $(f^1, f^2, m^{12}, u)$ as in the preceding example and for the two columns on the right: \\
\[               {\theta}^1_5 f^1 + {\theta}^2_5 f^2  + x^2  m^{12} + x^1 u + v^1  \,\,\,  ,   \hspace{1cm}    {\theta}^1_6 f^1 + {\theta}^2_6 f^2  - x^1 m^{12} + x^2 u  + v^2      \]
{\it a result which is totally new} because, though well known, {\it the explicit formulas for the infinitesimal generators of the elations have never been used in the literature up to our knowledge}.  \\
Using the fact that ${\xi}^2_{11} = - {\xi}^1_{12}$ and ${\xi}^2_{12} = {\xi}^1_{11}$, the integration by parts of the summation brings the following {\it totally new terms} after 
changing sign: \\
\[ A_1 =  {\xi}^1_{11} \longrightarrow {\partial}_r {\pi}^{1,r}  + {\mu}^{12,2} + {\nu}^1 = v^1  \,\, , \hspace{1cm} 
   A_2 = {\xi}^1_{12} \longrightarrow {\partial}_r {\pi}^{2,r}  -  {\mu}^{12,1} + {\nu}^2 = v^2  \]
We obtain therefore for the divergence corresponding to the left term:  \\
\[   \frac{1}{2} ((x^1)^2 - (x^2)^2) f^1 + x^1 x^2 f^2  + x^2 m^{12} + x^1 u + v^1 \hspace{7cm}  \]
\[ = \frac{1}{2} ((x^1)^2 - (x^2)^2) ({\partial_r {\sigma}^{1,r} ) +  x^1 x^2 ({\partial}_r { \sigma}^{2,r}) + x^2 ((\partial}_r{\mu}^{12,r} + {\sigma}^{2,1} - {\sigma}^{1,2})+ x^1 ( {\partial}_r  {\nu}^r     + {\sigma}^{1,1} + {\sigma}^{2,2})  + v^1           \]
\[ = ({\partial}_r ({\theta}^1_5 {\sigma}^{1,r}) - x^1 {\sigma}^{1,1} + x^2 {\sigma}^{1,2}) + {\partial}_r ({\theta}^2_5 {\sigma}^{2,r}) - (x^2 {\sigma}^{2,1} + x^1 {\sigma}^{2,2}) 
        + ( {\partial}_r (x^2 {\mu}^{12,r}) - {\mu}^{12,2} + x^2 {\sigma}^{2,1} - x^2 {\sigma}^{1,2} )        \]
\[        + ( {\partial}_r (x^1 {\nu}^r) - {\nu}^1  +  x^1 {\sigma}^{1,1} + x^1 {\sigma}^{2,2}) + v^1   \]
\[   =   {\partial}_r ({\theta}^i_5 {\sigma}^r_i   + x^2 {\mu}^{12,r}  + x^1 {\nu}^r + {\pi}^{1,r})     \hspace{10cm}              \]
Similarly, we also obtain for the right term:  \\
\[   x^1 x^2 f^1 + \frac{1}{2}((x^2)^2 - (x^1)^2) f^2  + x^1 m^{12} + x^2 u + v^2 \hspace{7cm}  \]
\[ = x^1 x^2 ({\partial_r {\sigma}^{1,r} ) + \frac{1}{2} ((x^2)^2 - (x^1)^2) ({\partial}_r { \sigma}^{2,r}) - x^1 ((\partial}_r{\mu}^{12,r} + {\sigma}^{2,1} - {\sigma}^{1,2})+ x^2 ( {\partial}_r  {\nu}^r     + {\sigma}^{1,1} + {\sigma}^{2,2})  + v^2           \]
\[ = ({\partial}_r ({\theta}^1_6 {\sigma}^{1,r}) - x^2 {\sigma}^{1,1} - x^1 {\sigma}^{1,2}) + ({\partial}_r ({\theta}^2_6 {\sigma}^{2,r}) - x^2 {\sigma}^{2,2} + x^1 {\sigma}^{2,1}) 
        -  {\partial}_r (x^1 {\mu}^{12,r}) + {\mu}^{12,1} - x^1 {\sigma}^{2,1} + x^1 {\sigma}^{1,2}         \]
\[        + ( {\partial}_r (x^2 {\nu}^r) - {\nu}^2  +  x^2 {\sigma}^{1,1} + x^2 {\sigma}^{2,2}) + v^2   \]
\[   =   {\partial}_r ({\theta}^i_6 {\sigma}^r_i  - x^1 {\mu}^{12,r}  + x^2 {\nu}^r + {\pi}^{2,r})     \hspace{10cm}              \]
Once again, we notice that the Spencer operator has been used in a crucial way and that no other classical approach could have allowed to obtain these results. \\  

We finally recall the fundamental diagram I and the corresponding adjoint diagram  that have never been used in mathematical physics:  \\

 \[  \fbox{  $ \begin{array}{rccccccccccccc}
 &&&&& 0 &&0&&0&  & \\
 &&&&& \downarrow && \downarrow && \downarrow  &\\
  & 0& \longrightarrow& \Theta &\stackrel{j_3}{\longrightarrow}& 6  &\stackrel{D_1}{\longrightarrow}&  12 &\stackrel{D_2}{\longrightarrow} &  6 &\longrightarrow  0 & \hspace{3mm}Spencer  \\
  &&&&& \downarrow & & \downarrow & & \downarrow & &    & \\
   & 0 & \longrightarrow &  2 & \underset 3{ \stackrel{j_3}{\longrightarrow}} & 20  & \stackrel{D_1}{\longrightarrow} &  
   30  &\stackrel{D_2}{\longrightarrow} & 12 &   \longrightarrow 0 &\\
   & & & \parallel && \hspace{5mm}\downarrow {\Phi}_0 & &\hspace{5mm} \downarrow {\Phi}_1 & & \hspace{5mm}\downarrow {\Phi}_2 &  &\\
   0 \longrightarrow & \Theta &\longrightarrow &  2 & \underset 3 {\stackrel{\cal{D}}{\longrightarrow}} &  14  & \stackrel{{\cal{D}}_1}{\longrightarrow} &  18  & \stackrel{{\cal{D}}_2}{\longrightarrow} &  6 & \longrightarrow  0 & \hspace{5mm} Janet       \\
   &&&&& \downarrow & & \downarrow & & \downarrow &      &\\
   &&&&& 0 && 0 && 0  &  &
   \end{array}   $  }  \]

 \[ \fbox{  $  \begin{array}{ccccccccccc}
   &&&& 0 &&0&&0&  & \\
   &&  && \uparrow && \uparrow && \uparrow  & &  \\
   & & 0 & \longleftarrow & 6  & \stackrel{ ad(D_1) }{\longleftarrow} &  12 & \stackrel{ ad(D_2) }{\longleftarrow} &  6 &   &  \\
   && \uparrow && \uparrow & & \uparrow & & \uparrow & &     \\
0 & \longleftarrow &  2 & \underset 3{ \stackrel{ ad(j_3) }{\longleftarrow} }& 20  & \stackrel{ ad(D_1) }{\longleftarrow} &  30  & \stackrel{ ad(D_2)}{ \longleftarrow} &  12 &  \longleftarrow  & 0  \\
   & & \parallel & & \uparrow  & & \uparrow  & & \uparrow &  &  \\
    0 & \longleftarrow &  2 & \underset 3{\stackrel{ ad(\cal{D}) }{\longleftarrow}} &  14  & \stackrel{ ad({\cal{D}}_1) }{\longleftarrow} &  \fbox{18}  & 
    \stackrel{ ad({\cal{D}}_2) }{\longleftarrow} &  6 &  \longleftarrow  & 0    \\
    & & \uparrow & & \uparrow  & & \uparrow  &  & \uparrow &    &   \\
    &  &  0  &  &  0  &  &  0  &  &  0 &  &  
   \end{array}   $  }  \]  \\

Three delicate snake chases are needed in order to prove that $ad({\cal{D}})$ is formally surjective, that the cohomology of the lower sequence at $\fbox{18}$ is isomorphic to the kernel of $ad(D_2)$ and finally that $ad({\cal{D}}_2)$ is formally injective, three facts that are not evident at first sight. As the second chase is not easy to work out, even for somebody familiar with homological algebra, we provide the details. For simplicity, starting with $a \in 18 $ means that $a$ is a section of the central vector bundle of fiber dimension $18$, that is locally a set of $18$ functions. If such an $a$ is killed by $ad({\cal{D}}_1)$, we can get $b\in 30$ killed by $ad(D_1)$ because the lower central square is commutative. As the central sequence is known to be locally exact [9, 13], then $b=ad(D_2) c$ for a certain $c \in 12$. Then we may project $c$ to $d \in 6$ which is killed by $ad(D_2)$ by the commutativity of the upper right square. Now, if $c\in 12$ should go to zero in $6$, then $c$ should come from $e \in 6$ and, applying $ad({\cal{D}}_2)$, we should obtain a coboundary in $18$. Finally, $ad({\cal{D}}_2)$ is injective because the central row of the diagram is a slitting sequence and $ad(D_2)$ is injective. We have the long $ker/coker $ exact sequence:  \\
\[     0 \rightarrow  D^2 \rightarrow D^{20} \rightarrow D^{30} \rightarrow D^{12} \rightarrow  0  \]
with the Euler-Poincar\'{e} formula: $ 12 - 30 + 20 - 2 = 0 $ in a coherent way.  \\

In the case of an arbitrary dimension $n\geq 1$, we have the commutative diagrams:  \\
\[ \fbox{  $    \begin{array}{lcccc} 
   &   & 0 &  & 0     \\
   &   & \downarrow &  &  \downarrow \\
       \lambda &  \in  & {\wedge}^0 T^* \otimes {\cal{G}} & \underset 1{ \stackrel{grad}{\longrightarrow}} & T^* \otimes  {\cal{G}} \\     
     \downarrow &  &  \downarrow     &    &  \downarrow         \\     
     {\xi}_q  & \in &  R_q  & \underset 1{ \stackrel{D_1}{\longrightarrow }}  & T^* \otimes R_q     \\ 
           &   & \uparrow &  &  \uparrow  \\
                  &   & 0 &  & 0                         
\end{array}  $  }   \]
and the corresponding adjoint diagram:  
\[  \fbox{  $         \begin{array}{ccccccc}
&  & 0  &  &  0  &  &  \\
 & &   \uparrow  &  &  \uparrow   & & \\
 0  & \longleftarrow & {\wedge}^n T^* \otimes  {\cal{G}}^* & \underset 1{\stackrel{div}{\longleftarrow}} &  {\wedge}^{n-1} T^* \otimes {\cal{G}}^* & \ni & {\cal{B}}    \\  
       &     &     \uparrow   &  & \uparrow  &  & \uparrow     \\
  0  &  \longleftarrow  &  {\wedge}^n T^* \otimes R^*_q  &  \underset 1{\stackrel{ad(D_1)}{\longleftarrow}} & {\wedge}^{n-1} T^* \otimes R^*_q   &  \ni  &  {\cal{X}}_q  \\
 &   & \uparrow  &  &  \uparrow  &  &  \\
   &  & 0  &  &  0 
\end{array}  $  }   \]

Accordingly, as "$e$" for "$elation$" is ranging from $1$ to $n$, setting ${\omega}_{es} x^s=x_e$, we are sure that:  \\
\[  {\partial}_r ({\theta}^i_e {\sigma}^r_i + a_e(x) \mu + x_e {\nu}^r + {\omega}_{se}{\pi}^{s,r} ) = {\theta}^i_e f_i + b_e(x) m + x_e u + {\omega}_{se} v^s  \]
with $a_e(x) $ and $b_e(x)$ are (tricky) linear combinations of $x$. For example, we have successively:  \\
\[  {\partial}_r ({\theta}^i_e {\sigma}^r_i) = {\theta}^i_e f_i + ({\partial}_r {\theta}^i_e) {\sigma}^r_i = {\theta}^i_e f_i  + x_e {\sigma}^r_r + ...  \]
\[  {\partial}_r( x_e {\nu}^r + {\omega}_{se}{\pi}^{s,r} ) = (x_e u - x_e {\sigma}^r_r + {\omega}_{se}{\nu}^s)+ {\omega}_{se} (v^s  - {\nu}^s ) + ... =  
x_e u - x_e {\sigma}^r_r  + {\omega}_{se} v^s + ...   \]
It follows that the trace of $\sigma$ disappears. The explicit computation of the remaining terms should be awfully technical, a fact explaining why these results are not known, even for $n=2$. The interest of using the Spencer sequence in the previous diagrams is to show out the interest of using "{\it sections}" of jet bundles instead of "{\it solutions}" but such an approach has never been done in mathematical physics. As a byproduct, we may quote:  \\

\noindent
{\bf THEOREM 2.5}: The general Maxwell/Weyl equations that are depending on the second order jets can be written as a pure divergence, totally independently of the fact that such a result is known in the framework discovered by Poincar\'{e} which is highly depending on the structure constants of the underlying Lie algebra as we already saw.  \\

\noindent
{\bf REMARK 2.6}: ({\it Special relativity}): Though surprising it may look like at first sight, the above example with $n=2$ perfectly fits with the original presentation of Lorentz transformations if one uses the "{\it hyperbolic}" notations $sh(\phi)=(e^{\phi} - e^{- \phi})/2, ch(\phi)= (e^{\phi}+ e^{-{\phi}})/2, th(\phi)= sh(\phi)/ch(\phi)$ with $ch^2(\phi) - sh^2(\phi) = 1$ instead of the classical $sin(\theta), cos (\theta), tan(\theta)= sin(\theta)/cos(\theta)$ with $cos^2(\theta) + sin^2(\theta) = 1$. Indeed, setting $x^1 = x, x^2 = c t$ and using the well defined  formula $th(\phi) = u/c$ among dimensionless quantities, the Lorentz transformation can be written:  
\[   {\bar{x}}^1 = \frac{x^1 - \frac{u}{c}x^2}{\sqrt{1- (\frac{u}{c})^2}}  ,  \,\,\,  {\bar{x}}^2 =  \frac{ - \frac{u}{c} x^1 + x^2}{\sqrt{1 - (\frac{u}{c})^2}}   \,\, \Leftrightarrow \,\,    
  {\bar{x}}^1 = ch(\phi) x^1 - sh(\phi) x^2, \,\,\,  {\bar{x}}^2 =  - sh(\phi) x^1  + ch(\phi) x^2        \]
Moreover, setting $th(\psi) = v/c$, we obtain easily for the composition of speeds:
\[ th(\phi + \psi) = \frac{(th(\phi) + th(\psi))}{( 1 + th(\phi) th(\psi))} \hspace{5mm} \Leftrightarrow  \hspace{5mm}
composition \,\, ( \frac{u}{c}, \frac{v}{c}) = \frac{(\frac{u}{c} + \frac{v}{c})}{(1 + \frac{u}{c}\frac{v}{c})}  \]  \\
without the need of any "{\it gedanken experiment}" on light signals.  \\

 \noindent
{\bf 3) ELECTROMAGNETISM AND GRAVITATION}  \\

When $n=4$, the comparison with the Maxwell equations of electromagnetism is easily obtained as follows. Indeed, writing a part of the dualizing summation in the form:\\
\[      {\cal{J}}^i ({\partial}_i A - A_i )+ \frac{1}{2} {\cal{F}}^{ij} ({\partial}_i A_j - {\partial}_j A_i) = -  {\cal{J}}^i A_i + {\sum}_{i\leq j} {\cal{F}}^{ij} ({\partial}_i A_j - {\partial}_j A_i) + ... \]
\[  = - {\cal{J}}^1A_1 + ... + {\cal{F}}^{12} ({\partial}_1 A_2 - {\partial}_2 A_1 + {\cal{F}}^{13} ({\partial}_1 A_3 - {\partial}_3 A_1) + {\cal{F}}^{14} ({\partial}_1 A_4 - {\partial}_4 A_1) + ...  \]
\[  =  - ({\cal{J}}^1 A_1+ ... + ({\cal{F}}^{12}{\partial}_2 A_1+  {\cal{F}}^{13}{\partial}_3 A_1 +  {\cal{F}}^{14}{\partial}_4 A_1) + ...) \]
\[   = div(...) + ( - {\cal{J}}^1 + {\partial}_2 {\cal{F}}^{12}+  {\partial}_3 {\cal{F}}^{13} +  {\partial}_4 {\cal{F}}^{14} ) A_1 + ... \]
Integrating by parts and changing the sign as usual, we obtain as usual the second set of Maxwell equations for the induction ${\cal{F}}$:  \\
\[    \fbox{ $  {\partial}_r {\cal{F}}^{ir} - {\cal{J}}^i =0  \,\,\, \Rightarrow  \,\,\, {\partial}_i {\cal{J}}^i = 0  $  }  \]
Such a result is coherent with the virial equation on the condition to have ${\sigma}^r_r = 0$ in a coherent way with the classical Maxwell impulsion-energy stress tensor density: \\
\[   {\sigma}^i_j =  {\cal{F}}^{ir} F_{rj} + \frac{1}{4} {\delta}^i_j {\cal{F}}^{rs} F_{rs}  \Rightarrow {\sigma}^r_r = 0   \]
which is traceless with a divergence producing the Lorentz force ([11] and [13] , p 444). Hence, the mathematical foundations of EM entirely depend on the structure of the conformal group of space-time, a fact that can only be understood by using the Spencer operator as we saw.  \\

As we have explained in the recent [31, 34], studying the mathematical structure of gravitation is much more delicate as it involves third order jets. Our purpose at the end of this paper is to consider only the linearized framework. The crucial idea is to notice that the Poisson equation has only to do with the trace of the stress tensor density, contrary to the EM situation as we just saw. \\

Defining the vector bundle ${\hat{F}}_0 = J_1(T)/{\hat{R}}_1 \simeq T^* \otimes {\hat{g}}_1$ when $n\geq 4$, another difficulty can be discovered in the following commutative and exact diagrams obtained by applying the Spencer $\delta$-map to the symbol sequence with $dim({\hat{g}}_1) = dim(g_1) + 1 = (n(n-1)/2 ) + 1$:
\[     0  \rightarrow  {\hat{g}}_1  \rightarrow   T^*\otimes T  \rightarrow   {\hat{F}}_0  \rightarrow  0   \]    
then to its first prolongation with $dim({\hat{g}}_2)= n$:   \\
\[  0   \rightarrow  {\hat{g}}_2 \rightarrow S_2T^*\otimes T  \rightarrow T^*\otimes {\hat{F}}_0 \rightarrow  0  \]
and finally to its second prolongation in which ${\hat{g}}_3= 0$:  \\
\[  \fbox{  $  \begin{array}{rcccccccccl}
  & &  0 & & 0 & & 0 &  &  &  & \\
  & & \downarrow & & \downarrow & & \downarrow & & & & \\
0 & \rightarrow & {\hat{g}}_3 & \rightarrow &  S_3T^*\otimes T & \rightarrow & S_2T^*\otimes {\hat{F}}_0& \rightarrow & {\hat{F}}_1 & \rightarrow & 0  \\
  & & \hspace{2mm}\downarrow  \delta  & & \hspace{2mm}\downarrow \delta & &\hspace{2mm} \downarrow \delta & & & & \\
0 & \rightarrow& T^*\otimes {\hat{g}}_2&\rightarrow &T^*\otimes S_2T^*\otimes T & \rightarrow &T^*\otimes T^*\otimes {\hat{F}}_0 &\rightarrow & 0 &&  \\
  & &\hspace{2mm} \downarrow \delta &  &\hspace{2mm} \downarrow \delta & &\hspace{2mm}\downarrow \delta &  &  & &  \\
0 & \rightarrow & {\wedge}^2T^*\otimes {\hat{g}}_1 & \rightarrow & \underline{ {\wedge}^2T^*\otimes T^*\otimes T }& \rightarrow & {\wedge}^2T^*\otimes {\hat{F}}_0 & \rightarrow & 0 &&  \\
 &  &\hspace{2mm}\downarrow \delta  &  & \hspace{2mm} \downarrow \delta  &  & \downarrow  & &  & & \\
0 & \rightarrow & {\wedge}^3T^*\otimes T & =  & {\wedge}^3T^*\otimes T  &\rightarrow   & 0  &  &  &  & \\
 &   &  \downarrow  &  &  \downarrow  &  &  &  &  &  &\\
  &  &  0  &   & 0  & &  &  &  &&
\end{array}  $  }  \]
A snake chase [37] allows to introduce the {\it Weyl} bundle ${\hat{F}}_1$ defined by the short exact sequence:  \\
\[      0  \longrightarrow T^* \otimes  {\hat{g}}_2 \stackrel{\delta}{\longrightarrow}  Z^2_1({\hat{g}}_1) \longrightarrow  {\hat{F}}_1 \longrightarrow 0 \]
in which the cocycle bundle $Z^2_1({\hat{g}}_1)$ is defined by the short exact sequence: \\
\[   0 \longrightarrow Z^2_1({\hat{g}}_1) \longrightarrow  {\wedge}^2 T^* \otimes {\hat{g}}_1 \stackrel{\delta}{\longrightarrow} {\wedge}^3 T^* \otimes T \longrightarrow 0  \]
We have of course $dim({\hat{F}}_1)= 10$ when $n=4$ but more generally: \\
\[   \begin{array}{ccl}
dim({\hat{F}}_1) &  =  &  (n (n+1)/2)(n(n+1)/2 - 1) - n^2(n + 1)(n + 2)/6 \\
                           &  =  &  ((n(n-1)/2)(n(n-1)/2 + 1) - n^2 (n-1)(n-2)/6) - n^2  \\ 
                           &  =  &  n(n+1)(n+2)(n-3)/12
\end{array}   \]
In the purely Riemannian case, as $g_2=0$, we have $ F_1 \simeq Z^2_1(g_1)$ and thus:  \\
\[   \begin{array}{ccl}
dim(F_1) &  =  &  (n (n+1)/2)(n(n+1)/2 - 1) - n^2(n + 1)(n + 2)/6 \\
                           &  =  &  n^2 (n-1)/2)^2- n^2 (n-1)(n-2)/6    \\
                           &  =  &   n^2 (n^2 - 1)/12 
\end{array}   \]
a result leading to the unexpected formula $dim(F_1) - dim({\hat{F}}_1) = n(n + 1) / 2$. \\

Needless to say that no classical method can produce such results which are summarized in the following {\it Fundamental Diagram II} provided as early as in $1983$ [10, 12]:  \\ 
\[  \fbox{  $   \begin{array}{rcccccccccl}
&  &  &  &  &  &  &  &  &  & \\
  &  &  &  &  &  &   &  &  0  & & \\
  &  &  &  &  &  &   &  & \downarrow &  & \\
  &  &  &  &  &  &  0  &  &  Ricci & &  \\
  &  &  &  &  &  &  \downarrow &  & \downarrow &  &  \\
  &  &  &  &  0 & \longrightarrow & Z^2_1(g_1) & \longrightarrow & Riemann & \longrightarrow & 0  \\
  &  &  &  &   \downarrow &  & \downarrow &  & \downarrow &  &  \\
  &  & 0 & \longrightarrow & T^* \otimes {\hat{g}}_2 & \stackrel{\delta}{\longrightarrow} & Z^2_1({\hat{g}}_1) & \longrightarrow & Weyl  & \longrightarrow & 0  \\
  &  &  &  &  \downarrow &  &  \downarrow &  &  \downarrow &  &  \\
 0 &  \longrightarrow & S_2T^* &  \stackrel{\delta}{\longrightarrow} &  T^* \otimes T^* & \stackrel{\delta}{\longrightarrow} & {\wedge}^2 T^* &  \longrightarrow & 0 &  &  \\      
  &  &  &  & \downarrow &  &  \downarrow &  &  &  &  \\
  &  &  &  &  0 &  &  0  &  &  &  &\\
  &  &  &  &  &  &  &  &  &  & 
  \end{array}  $  }  \]   \\

\noindent
{\bf THEOREM 3.1}: This commutative and exact diagram splits and a diagonal snake chase proves that $Ricci \simeq S_2T^*$.  \\

\noindent
{\it Proof}: The monomorphism $\delta: S_2 T^* \rightarrow T^* \otimes T^*$ splits with $\frac{1}{2} (A_{i,j} + A_{j,i}) \leftarrow A_{i,j}$ while the epimorphism $\delta: T^* \otimes T^* \rightarrow {\wedge}^2 T^*:A_{i,j} \rightarrow A_{i,j} - A_{j,i}$ splits with $ \frac{1}{2} F_{ij} \leftarrow F_{ij}= - F_{ji}$. We explain therefore how the well known result $T^* \otimes T^* \simeq S_2T^* \oplus {\wedge}^2 T^*$, which is coming from the elementary formula $n^2 = n(n+1)/2 + n(n-1)/2$ may be related to the Spencer $\delta$-cohomology interpretation of the Riemann and Weyl bundles. For this, we have to give details on the chase:  \\
\[  S_2T^* \rightarrow T^* \otimes T^* \rightarrow T^* \otimes {\hat{g}}_2 \rightarrow Z^2_1({\hat{g}}_1) \rightarrow Z^2_1(g_1) \rightarrow Riemann \rightarrow Ricci  \]
Starting with $(A_{ij} = A_{i,j} = A_{j,i} = A_{ji} ) \in S_2T^*\subset T^* \otimes T^* $, we may define:  \\
\[  {\xi}^r_{ri,j} = n A_{i,j}=n A_{ij} = nA_{ji} = {\xi}^r_{rj,i} \Rightarrow ({\xi}^k_{lj,i} ={\delta}^k_l A_{j,i} + {\delta}^k_j A_{l,i} - {\omega}_{ij} {\omega}^{kr} A_{r,i}) \in T^* \otimes {\hat{g}}_2  \]           
\[  \Rightarrow  (R^k_{l,ij} = {\xi}^k_{li,j} - {\xi}^k_{lj,i} ) \in Z^2_1({\hat{g}}_1) \in {\wedge}^2 T^* \otimes {\hat{g}}_1 \in {\wedge}^2 T^* \otimes T^* \otimes T \]
\[ \Rightarrow   R^r_{r,ij} = {\xi}^r_{zi,j} - {\xi}^r_{rj,i} = n (A_{i,j} - A_{j,i}) = 0 \Rightarrow (R^k_{l,ij}) \in   Z^2_1 (g_1)\]
\[  \Rightarrow  n R^k_{l,ij} = ({\delta}^k_l {\xi}^r_{ri,j} +{\delta}^k_i {\xi}^r_{rl,j}  - {\omega}_{li} {\omega}^{ks} {\xi}^r_{rs,i}) - ( {\delta}^k_l{\xi}^r_{rj,i} + {\delta}^k_j {\xi}^r_{rl,i} - 
{\omega}_{lj} {\omega}^{ks} {\xi} ^r_{rs,i})  \]                                                                                                                                                                                                                                                  
\[  \Rightarrow  R^k_{l,ij} = ({\delta}^k_i A_{lj} - {\delta}^k_jA_{li}) - {\omega}^{ks} ( {\omega}_{li} A_{sj} - {\omega}_{ lj} A_{si} )  \]
Introducing $tr(A) = {\omega}^{ij}A_{ij} $ and  $R_{ij} = R^r_{ i,rj} = (n A_{ij} - A_{ij} ) - (A_{ij} - {\omega}_{lj} tr(A))  $, we get:  \\                                                                                                                                                                                                                                                                                                                                                                                                                                                           
\[  \fbox{  $  R_{ij} = (n - 2) A_{ij}  + {\omega}_{ij} tr(A) = R_{ji}  \Rightarrow  tr(R) = {\omega}^{ij} R_{ij} = 2 ( n - 1) tr(A) $  } \]
Substituting, we finally obtain $A_{ij} = \frac{1}{(n-2)} R_{ij} - \frac{1}{ 2 (n-1)(n-2)} {\omega}_{ij} tr (R)$ and the tricky formula:    \\                                   
\[ \fbox{  $   R^k_{l,ij} = \frac{1}{(n-2)} ({\delta}^k_i R_{lj} - {\delta}^k_j R_{li}  - {\omega}^{ks} ({\omega}_{li} R_{sj}- {\omega}_{lj} R_{si}) - 
\frac{1}{(n-1)(n-2)} ({\delta}^k_i {\omega}_{lj} - {\delta}^k_j {\omega}_{li} ) tr(R)  $  }   \]                                           
totally independently from the standard elimination of the derivatives of a conformal factor.  \\
Contracting in $k$ and $i$, we obtain indeed the lift:  \\
\[    Riemann = H^2_1(g_1) \rightarrow S_2 T^* \simeq Ricci: R^k_{l,ij} \rightarrow R^r_{i,rj} = R_{ij} = R_{ji}  \]
in a coherent way. Using a standard result of homological algebra [16, 17, 37], we obtain therefore a splitting
  $Weyl = H^2_1({\hat{g}}_1) \rightarrow H^2_1(g_1) = Riemann $:  \\
\[ \fbox{  $  W^k_{l,ij} = R^k_{l,ij}  -  ( \frac{1}{(n-2)} ({\delta}^k_i R_{lj} - {\delta}^k_j R_{li}  - {\omega}^{ks} ({\omega}_{li} R_{sj}- {\omega}_{lj} R_{si}) - 
\frac{1}{(n-1)(n-2)} ({\delta}^k_i {\omega}_{lj} - {\delta}^k_j {\omega}_{li} ) tr(R))  $  }   \]   
in such a way that $W^r_{i,rj} = 0$, a result leading to the isomorphism $Riemann \simeq Ricci \oplus Weyl$.   \\
\hspace*{12cm}  $\Box$  \\

Introducing the Christoffel symbols ${\gamma}^k_{ij}$ of the metric $\omega$, we may define the (true) $1$-form [13] : 
\[   {\alpha}_i = {\chi}^r_{r,i} + {\gamma}^r_{r,s} {\chi}^s_{,i} = n({\partial}_i a - A^r_i a_r) = n({\partial}_i a - a_i) - na_r g^r_s ({\partial}_i f^s - f^s_i)  \]
because we have $\gamma = 0$ in the conformal case and, taking the respective determinants:
\[ {\omega}_{kl}(f(x)) f^k_i (x) f^l_j (x)= e^{2 \, a(x)} {\omega}_{ij}(x)  \Rightarrow   det(f^l_j) = e^{na}  \Rightarrow  g^r_l {\partial}_i f^l_r = \frac{1}{det(f^l_j)} {\partial}_i det(f^l_j)= 
n{\partial}_i a  \]
It follows that:
\[ \left\{ \begin{array}{lcl}
 {\Omega}_{ij} & \equiv & (L({\hat{\xi}}_1)\, \omega)_{ij}  = 2 A \,  {\omega}_{ij} \\  
  {\Gamma}^k_{ij} & \equiv  &  (L({\hat{\xi}}_2 )\,  \gamma ) ^k_{ij} = {\delta}^k_i A_j + {\delta}^k_j A_i - {\omega}_{ij}  {\omega}^{kr}A_r                                     
 \end{array}    \right.         \]
after linearization, in such a way that ${\partial}_i a - a_i = 0  \Rightarrow  {\partial}_i A - A_i = 0$ for a formally integrable system of defining finite Lie equations. It is important to notice that the geometric objects appearing in the Janet sequence, namely $(\omega, \gamma) \simeq j_1(\omega)$ according to the Levi-Civita isomorphism, are quite different from the ${\chi}_q$ appearing in the Spencer sequence that have never been introduced in mathematical physics and have only been introduced for the first time in the study of Cosserat media [3, 14, 15].  \\

We provide a short unusual motivating example in order to convince the reader that the construction of differential sequences is not an easy task indeed.  \\

\noindent

\noindent
{\bf EXAMPLE 3.2}: (Macaulay example) \\
With $m = 1, n = 3, q = 2, K = \mathbb{Q}$, let us consider the linear second order system $R_2 \subset J_2(E)$ with $dim(E) = 1$ provided by F. S. Macaulay in $1916$ [66, 25, 26] while using jet notations: 
\[  y_{33} = 0, \hspace{1cm}  y_{23} - y_{11} = 0, \hspace{1cm} y_{22} = 0  \]
We let the reader check easily that $dim(g_2)= 3, dim(g_3) = 1$ with only parametric jet $y_{111}$, $g_4 = 0 $ and thus $dim( R_2)= 8 = 2^3$, a result leding to s $dim(R_{3 + r}) = 8$ that is $R_{3+r}\simeq R_3, \forall r\geq 0$.  We recall the dimensions of the following jet bundles:  
\[  \begin{array}{cccccccccc}
q & \rightarrow & 0 & 1 & 2 & 3 & 4 & 5 & 6 & 7   \\
S_qT^* & \rightarrow & 1 & 3 & 6 & 10 & 15 & 21 & 28 & 36  \\
J_q(E) & \rightarrow & 1 & 4 & 10 & 20 & 35 & 56 & 84 & 120
\end{array}  \]
and the generic commutative and exact diagram allowing to construct the Spencer bundles $C_r \subset C_r (E)$ and the Janet bundles $F_r$ for $r= 0, 1, ... ,n$ with $F_0=J_q(E)/R_q$ from the short exact sequence of vector bundles:
\[    0 \rightarrow g_{q+1} \rightarrow S_{q+1} T^* \otimes E \rightarrow h_1 \rightarrow 0 \,\,\, \Rightarrow 
\,\,\, h_1 \subset T^*\otimes F_0 \]
\[ \fbox{  $  \begin{array}{ccccccc}
0 &  & 0 & & 0 & & \\
\downarrow & & \downarrow & & \downarrow & & \\
{\wedge}^{r-1}T^* \otimes g_{q+1} & \stackrel{\delta} { \longrightarrow} & \underline{{\wedge}^r T^* \otimes R_q } & \longrightarrow & C_r & \longrightarrow & 0  \\
\downarrow &  &  \downarrow &  &  \downarrow &  &  \\
\underline{{\wedge}^{r-1}T^* \otimes S_{q+1} T^* \otimes E }  & \stackrel{\delta}{\longrightarrow} & \underline{{\wedge}^r T^* \otimes J_q(E) }& \longrightarrow & C_r(E) & \longrightarrow & 0  \\
\downarrow &  &  \downarrow & \searrow  &  \downarrow  &    &   \\
{\wedge}^{r-1}T^* \otimes h_1 & \stackrel{\delta}{\longrightarrow} & {\wedge}^r T^* \otimes F_0 & \longrightarrow & F_r & \rightarrow & 0   \\
\downarrow &  &  \downarrow &  &  \downarrow   &  &  \\
   0  &  &  0  &  &  0  &  &  
\end{array}   $  }  \]
showing that we have indeed:   \\
\[    C_r = {\wedge}^r T^* \otimes R_q / \delta ({\wedge}^{r-1} T^* \otimes g_{q+1}) \]
\[    C_r(E) = {\wedge}^r T^* \otimes J_q(E) / \delta ({\wedge}^{r-1} T^* \otimes S_{q+1}T^* \otimes E)    \]
\[   F_r = {\wedge}^r T^* \otimes J_q(E) / ( {\wedge}^r T^* \otimes R_q + \delta ({\wedge}^{r-1} T^* \otimes S_{q+1}T^ \otimes E))  \]
When $R_q\subset J_q(E)$ is involutive, that is formally integrable (FI) with an involutive symbol $g_q$, then these three differential sequences are formally exact on the jet level and, in the Spencer sequence: \\
\[ 0 \longrightarrow  \Theta \stackrel{j_q}{\longrightarrow} C_0 \stackrel{D_1}{\longrightarrow} C_1 \stackrel{D_2}{\longrightarrow} ... \stackrel{D_n}{\longrightarrow} C_n 
\longrightarrow 0  \]
the first order involutive operators $D_1, D_2, ..., D_n$ are induced by the Spencer operator $d:R_{q+1} \rightarrow T^* \otimes R_q$ already considered that can be extended to 
$d: {\wedge}^r T^* \otimes R_{q+1}\rightarrow {\wedge}^{r+1} \otimes R_q$. A similar condition is also valid for the Janet sequence:  \\
\[  0 \longrightarrow \Theta \longrightarrow E \stackrel{{\cal{D}}}{\longrightarrow} F_0 \stackrel{{\cal{D}}_1}{\longrightarrow} F_1 \stackrel{{\cal{D}}_2}{\longrightarrow} ... \stackrel{{\cal{D}}_n}{\longrightarrow} F_n \longrightarrow 0  \]
which can be thus constructed "as a whole" from the previous extension of the Spencer operator (See [9] p 183 + 185 + 391 for the main diagrams and [25] for other explicit computations on the Macaulay example). However, this result is still not known and not even acknowledged today in mathematical physics, particularly in general relativity which is {\it never} using the Spencer $\delta$-cohomology in order to define the Riemann or Bianchi operators [20, 24, 27]. The study of the present Macaulay example will be sufficient in order to justify our comment.  \\
First of all, as $g_2$ is {\it not} $2$-acyclic and the coeficients are constant, the CC are of order two as follows:\\
\[ Q w - R v = 0,  \,  R u - P w = 0, \, P v - Q u =  0 \, \Rightarrow \,  P (Q w - R v ) + Q (R u - P w ) + R (P v - Q u) \equiv  0    \]  
The simplest formally exact resolution, {\it which is quite far from being a Janet sequence}, is thus: \\
\[   \fbox{  $  0  \longrightarrow \Theta \longrightarrow 1 \underset 2{\stackrel{\cal{D}}{\longrightarrow}}  3  
\underset 2{\stackrel{{\cal{D}}_1}{ \longrightarrow }} 3 \underset 2{\stackrel{{\cal{D}}_2}{\longrightarrow }}
1 \longrightarrow 0  $  }  \]                                       
Secondly, as the first prolongation of $R_2$ becoming involutive is $R_4$, an idea could be to start with the system $R_3 \subset J_3(E)$ but we have proved in [LAP] that the simplest formally exact sequence that could be obtained, {\it which is also quite far from being a Janet sequence}, is:  \\
\[  \fbox{ $    0 \longrightarrow \Theta \longrightarrow 1\underset 3{\longrightarrow} 12 \underset 1{\longrightarrow} 21 \underset 2{\longrightarrow} 46 \underset 1{\longrightarrow} 72 
\underset 1{\longrightarrow} 48 \underset 1{\longrightarrow} 12 \longrightarrow 0   $  }  \]
Indeed, the Euler-Poincar\'{e} characteristic is $1 - 12 + 21 - 46 + 72 - 48 + 12 = 0 $ but  we notice that the orders of the successive operators may vary up and down. \\

We finally let the reader discover that the {\it Fundamental Diagram I} relating the upper Spencer sequence to the lower Janet sequence is (See [23], p 19-22 for details):  \\
\[  \fbox{  $    \begin{array}{ccccccccccccccc}
 &  &  &  &  &  &  0 &  &  0  &  &  0  &  &  0  &  &   \\
&  &  &  &  &  & \downarrow &  &  \downarrow  &  &  \downarrow &  & \downarrow &  &  \\
  &  &  0  & \longrightarrow & \Theta & \underset 4{\stackrel{j_4}{\longrightarrow}} & 8 & \underset 1{\stackrel{D_1}{\longrightarrow}} & 24 & \underset 1{\stackrel{D_2}{\longrightarrow}} & 24 & \underset 1{\stackrel{D_3}{\longrightarrow}} & 8 & \longrightarrow & 0  \\
&  &  &  &  &  & \downarrow &  &  \downarrow  &  &  \downarrow &  & \downarrow &  &  \\
 &  &  0  & \longrightarrow & 1 & \underset 4{\stackrel{j_4}{\longrightarrow}} & 35 & \underset 1{\stackrel{D_1}{\longrightarrow}} & 84 & \underset 1{\stackrel{D_2}{\longrightarrow}} & 70 & \underset 1{\stackrel{D_3}{\longrightarrow}} & 20 & \longrightarrow & 0  \\
&  &  &  &  \parallel &  & \downarrow &  &  \downarrow  &  &  \downarrow &  & \downarrow &  &  \\
0 & \longrightarrow & \Theta & \longrightarrow & 1 & \underset 4{\stackrel{{\cal{D}}}{\longrightarrow}} & 27 & \underset 1{\stackrel{{\cal{D}}_1}{\longrightarrow}} & 60 & \underset 1{\stackrel{{\cal{D}}_2}{\longrightarrow}} & 46 & \underset 1{\stackrel{{\cal{D}}_3}{\longrightarrow}} &  12 & \longrightarrow & 0\\ 
&  &  &  &  &  & \downarrow &  &  \downarrow  &  &  \downarrow &  & \downarrow &  &  \\
 &  &  &  &  &  &  0 &  &  0  &  &  0  &  &  0  &  &
\end{array}   $  }   \]
In the present example, the Spencer bundles are $C_r ={\wedge}^r T^* \otimes R_4$ and their dimensions are quite lower that the dimensions of the Janet bundles. Among the long exact sequences that {\it must} be used the following involves a $540 \times 600$ matrix and we wish good luck to anybody using computer algebra:  \\
\[   0 \rightarrow R_7 \rightarrow J_7(1) \rightarrow J_3(27) \rightarrow J_2(60) \rightarrow J_1(46) \rightarrow F_3 \rightarrow 0  \]
\[  \Rightarrow \,\,\,  8 - 120 + 540 - 600 + 184 = 12 = dim(F_3) \]

We are now ready to apply the previous diagrams by proving the following crucial Theorem:  \\

\noindent
{\bf THEOREM  3.3}: When $n=4$, the linear Spencer sequence for the Lie algebra ${\hat{\Theta}}$  of infinitesimal conformal group of transformations projects onto a part of the Poincar\'{e} sequence for the exterior derivative {\it with a shift by one step} according to the following commutative and locally exact diagram:  
\[ \fbox{  $   \begin{array}{rcccccccl}
0 \longrightarrow & \hat{\Theta} & \stackrel{j_2}{\longrightarrow} & {\hat{R}}_2 & \stackrel{D_1}{\longrightarrow} & T^* \otimes  {\hat{R}}_2 & \stackrel{D_2}{\longrightarrow} &  {\wedge}^2T^*\otimes {\hat{R}}_2   \\
&  &  &  \downarrow & \swarrow &  \downarrow  &  &  \downarrow   \\
&  &  &  T^* & \stackrel{d}{\longrightarrow}  & {\wedge}^2T^* & \stackrel{d}{\longrightarrow} &  {\wedge}^3T^* \\
&  &  &  A &  &  d A=F &  &d F=0   
\end{array}   $  }  \]
{\it This purely mathematical result also contradicts classical gauge theory} because it proves that EM only depends on the structure of the conformal group of space-time but not on $U(1)$.  \\

\noindent
{\it Proof}: Considering $\omega$ and $\gamma$ as geometric objects, we obtain at once the formulas:\\
\[{\bar{\omega}}_{ij}=e^{2a(x)}{\omega}_{ij} \hspace{5mm}  \Rightarrow  \hspace{5mm}  {\bar{\gamma}}^r_{ri}={\gamma}^r_{ri} + 
{\partial}_ia  \]
With more details, on the infinitesimal level, we have successively:  
\[ L({\xi}_1)\omega=2A\omega \Rightarrow  ({\xi}^r_r+{\gamma}^r_{ri}{\xi}^i)=nA,\forall {\hat{\xi}}_1 \in {\hat{R}}_1  \]
\[  (L({\xi}_2)\gamma)^k_{ij}={\delta}^i A_j + {\delta}^k_j A_i -  {\omega}_{ij} {\omega}^{kr} A_r \Rightarrow 
{\xi}^r_{ri} + {\gamma}^r_{rs}{\xi}^s_i + {\xi}^s {\partial}_s{\gamma}^r_{ri} = n A_i, 
 \forall {\xi}_2\in {\hat{R}}_2\]

We obtain therefore an isomorphism $J_1({\wedge}^0T^*)\simeq {\wedge}^0T^*{\times}_XT^*$, a result leading to the following commutative diagram: \\
\[  \begin{array}{rcccccl}
0 \longrightarrow & R_2 & \longrightarrow & {\hat{R}}_2 & \longrightarrow  & J_1({\wedge}^0T^*) & \longrightarrow 0  \\
  & \hspace{3mm}\downarrow d  &  & \hspace{3mm}  \downarrow d  & &\hspace{3mm} \downarrow d  &   \\
0 \longrightarrow & T^* \otimes R_1 & \longrightarrow & T^* \otimes {\hat{R}}_1 & \longrightarrow  & T^* & \longrightarrow 0     
\end{array}    \]
where the rows are exact by counting the dimensions. The operator $d:(A,A_i) \longrightarrow ({\partial}_iA-A_i)$ on the right is induced by the central Spencer operator, a result that could not have been even imagined by Weyl and followers. This result provides a good transition towards the conformal origin of electromagnetism. \\
We now restrict our study to the linear framework and introduce a new system ${\tilde{R}}_1\subset J_1(T)$ of infinitesimal Lie equations defined by $L({\xi}_1) \omega = 2 A \omega$ with prolongation                                                                                                                                                                                                                                                                                                                                                                                                                                                                                                                                                                                                                                                              defined by $L({\xi}_2)\gamma = 0$  in such a way that 
$R_1 \subset   {\tilde{R}}_1 = {\hat{R}}_1$ with a strict inclusion and the strict inclusions 
$R_2 \subset {\tilde{R}}_2 \subset {\hat{R}}_2$. \\

\noindent
{\bf LEMMA 3.4}: One has an isomorphism ${\hat{R}}_2 / {\tilde{R}}_2 \simeq {\hat{g}}_2$.  \\

\noindent
{\it Proof}: From the definitions, we obtain the following commutative and exact diagram:  \\
\[       \begin{array}{ccccccccc}
  &  &  &  &  0 &  &  &  & \\
  &  &  &  & \downarrow & &  &  &  \\
  &  &  0 & \rightarrow & {\hat{g}}_2  &  &  &  &  \\  
  &  &  \downarrow &  & \downarrow & \searrow &  &  &  \\
  0 & \rightarrow & {\tilde{R}}_2 & \rightarrow & {\hat{R}}_2 & \rightarrow& {\hat{R}}_2/{\tilde{R}}_2 & \rightarrow & 0 \\
  &  &  \downarrow & & \downarrow & & \downarrow &   &   \\
  0 & \rightarrow & {\tilde{R}}_1 & =  & {\hat{R}}_1&  \rightarrow &  0 &  &  \\
  &  &  \downarrow  &  &  \downarrow &  &  &  &  \\
  &  &  0  &  &  0  &  &  &  & 
  \end{array}   \]
The  south-east arrow is an isomorphism as it is both a monomorphism and an epimorphism by using a snake chase showing that ${\hat{R}}_2 = {\tilde{R}}_2 \oplus {\hat{g}}_2$.  \\
\hspace*{13cm}  $\Box$

A first problem to solve is to construct vector bundles from the components of the image of $D_1$. Using the corresponding capital letter for denoting the linearization, let us introduce:   \\
\[   {\partial}_i {\xi}^k_{\mu} - {\xi}^k_{\mu + 1_i} = X^k_{\mu, i} \,\,  \Rightarrow \,\,  B^k_{\mu, i} \,\, (tensors) \]
\[ (B^k_{l,i}=X^k_{l,i}+{\gamma}^k_{ls}X^s_{,i}) \in T^*\otimes T^*\otimes T \Rightarrow  (B^r_{r,i}=B_i)\in T^*\]
\[  (B^k_{lj,i}=X^k_{lj,i}+{\gamma}^k_{sj}X^s_{l,i}+{\gamma}^k_{ls}X^s_{j,i}-{\gamma}^s_{lj}X^k_{s,i}+X^r_{,i}{\partial}_r{\gamma}^k_{lj}) \in T^*\otimes S_2T^*\otimes T \Rightarrow (B^r_{ri,j}-B^r_{rj,i}=F_{ij})\in {\wedge}^2T^*  \]
We obtain from the relations ${\partial}_i{\gamma}^r_{rj}={\partial}_j{\gamma}^r_{ri}$ and the previous results:  \\
\[ \begin{array}{rcl}
F_{ij}=B^r_{ri,j}-B^r_{rj,i} & = & X^r_{ri,j}-X^r_{rj,i}+{\gamma}^r_{rs}X^s_{i,j}-{\gamma}^r_{rs}X^s_{j,i}+X^r_{,j}{\partial}_r{\gamma}^s_{si}-X^r_{,i}{\partial}_r{\gamma}^s_{sj}  \\
  &  =  & {\partial}_iX^r_{r,j}-{\partial}_jX^r_{r,i}+{\gamma}^r_{rs}(X^s_{i,j}-X^s_{j,i})+X^r_{,j}{\partial}_i{\gamma}^s_{sr}-X^r_{,i}{\partial}_j{\gamma}^s_{sr} \\
    &  =  & {\partial}_i(X^r_{r,j}+{\gamma}^r_{rs}X^s_{,j})-{\partial}_j(X^r_{r,i}+{\gamma}^r_{rs}X^s_{s,i})  \\
      &  =  &  {\partial}_iB_j-{\partial}_jB_i
      \end{array}   \]
Now, using the contracted formula ${\xi}^r_{ri}+ {\gamma}^r_{rs}{\xi}^s_i + {\xi}^s{\partial}_s{\gamma}^r_{ri}=nA_i$, we obtain:  \\
\[ \begin{array}{rcl}
 B_i & =  & ({\partial}_i{\xi}^r_r - {\xi}^r_{ri})+{\gamma}^r_{rs}({\partial}_i{\xi}^s - {\xi}^s_i)\\
    &  =  &{\partial}_i{\xi}^r_r + {\gamma}^r_{rs}{\partial}_i{\xi}^s+
{\xi}^s {\partial}_s{\gamma}^r_{ri} - nA_i \\
  &  =  &{\partial}_i({\xi}^r_r + {\gamma}^r_{rs}{\xi}^s) - nA_i \\
    &  =  &n ({\partial}_iA - A_i) 
\end{array}   \]  
and we finally get $F_{ij}=n({\partial}_jA_i-{\partial}_iA_j)$ {\it which is no longer depending on} $A$, a result fully solving the dream of Weyl. Of course, when $n=4$ and $\omega$ is the Minkowski metric, then we have $\gamma=0$ in actual practice and the previous formulas become particularly simple. \\     

It follows that $d B=F \Leftrightarrow - ndA=F$ in ${\wedge}^2T^*$ and thus $d F=0$, that is $ {\partial}_i{F}_{jk} + {\partial}_j{F}_{ki} + {\partial}_k{F}_{ij}=0 $, has an intrinsic meaning in ${\wedge}^3T^*$. It is finally important to notice that the usual EM Lagrangian is defined on sections of ${\hat{C}}_1$ killed by $D_2$ but {\it not} on ${\hat{C}}_2$. Finally, the south west arrow in the left square is the composition:   \\
\[  {\xi}_2 \in {\hat{R}}_2 \stackrel{D_1}{\longrightarrow} X_2 \in T^* \otimes {\hat{R}}_2 \stackrel{{\pi}^2_1}{\longrightarrow } X_1 \in T^*\otimes {\hat{R}}_1 \stackrel{(\gamma)}{\longrightarrow} (B_i) \in  T^*  \Leftrightarrow {\xi}_2 \in {\hat{R}}_2 \rightarrow (nA_i)\in T^*  \]
More generally, using the Lemma, we have the composition of epimorphisms:  \\
\[  {\hat{C}}_r \rightarrow {\hat{C}}_r/ {\tilde{C}}_r \simeq {\wedge}^r T^* \otimes ({\hat{R}}_2 / {\tilde{R}}_2) \simeq {\wedge}^r T^* \otimes {\hat{g}}_2 \simeq {\wedge}^r T^* \otimes T^* \stackrel{\delta}{\rightarrow} {\wedge}^{r+1} T^* \]
Accordingly, though $A$ and $B$ are potentials for $F$, then $B$ can also be considered as a part of the {\it field} but the important fact is that the first set of ({\it linear}) Maxwell equations $d F=0$ is induced by the ({\it linear}) operator $D_2$ because we are only dealing with involutive and thus formally integrable operators, a {\it fact} justifying the commutativity of the square on the left of the diagram.   \\
\hspace*{12cm}     $\Box $     \\

\noindent
{\bf REMARK 3.5}: Taking the determinant of each term of the non-linear second order PD equations defining $\hat{\Gamma}$ while using jet notations, we obtain successively:  \\
\[    {\omega}_{kl}(y)y^k_i y^l_j = e^{2a(x)} {\omega}_{ij}(x) \Rightarrow det(\omega)(det(f^k_i)^2  = e^{2na(x)} det(\omega) \Rightarrow det(y^k_i)= e^{na(x)} \]  
in such a way that we may define $b(f(x))= a(x)$ over the target when $\Delta (x) = det({\partial}_i f^k(x)) \neq 0$ while caring only about the connected component $0[ \rightarrow 1 \rightarrow \infty$ of the dilatation group. The problem is thus to pass from a (metric) tensor to a (metric) tensor density and to consider successively the two non-linear systems of finite defining Lie equations:   \\
\[  {\omega}_{kl}(y)y^k_i y^l_j ={\omega}_{ij}(x) \,\,\, \rightarrow \,\,\, {\hat{\omega}}_{kl}y^k_i y^l_j (det(y^k_i))^{\frac{-2}{n}}={\hat{\omega}}_{ij}(x) \]
 Now, with $\gamma=0$ we have ${\chi}^r_{r,i} = g^s_k ( {\partial}_i f^k_s - A^r_if^k_{rs})$ and:   \\
 \[    g^s_k{\partial}_i f^k_s= (1/det(f^k_i)){\partial}_i det(f^k_i) =n{\partial}_ia , \,\,\,   g^s_k f^k_{rs}= na_r(x)  \]
 Finally, we have the jet compositions and contractions:  \\
\[   g^r_kf^k_i={\delta}^r_i \Rightarrow g^r_kf^k_{ij}= - g^r_{kl}f^k_if^l_j \,\, \Rightarrow \,\,  n\, a_i(x)=g^s_kf^k_{is}= - f^k_if^l_r g^r_{kl}= 
- n\, f^k_i(x)b_k(f(x))  \] 
It follows that ${\alpha}_i= n({\partial}_i a(x) - A^r_ia_r(x))$ but we may also set $a_i=f^k_ib_k$ in order to obtain simply ${\alpha}_i=  n ( \frac{\partial b}{\partial y^k} - b_k) {\partial}_if^k$ as a way to pass from source to target (Compare to [28]). \\   \\  \\

We just sketch now the nonlinear framework and a few among its consequences [12, 13, 19].  \\

When $X$ is a manifold of dimension $n$ and $Y$ is just a copy of $X$, let us consider the $q$-jet bundle ${\Pi}_q = {\Pi}_q(X, Y)$ of invertible jets of order $q$ with local coordinates 
$(x, y_q) = (x^i, y^k, y^k_i, y^k_{ij}, ...)$ such that $det(y^k_i)\neq 0$. There is a canonical embedding ${\Pi}_{q+1} \subset J_1({\Pi}_q)$ defined by $y^k_{\mu,i} = y^k_{\mu + 1_i}$.
If ${\cal{R}}_q \subset {\Pi}_q$ is a Lie groupoid of order $q$ with first prolongation ${\cal{R}}_{q+1} = J_1({\cal{R}}_q) \cap {\Pi}_{q+1} \subset J_1 ({\Pi}_q)$ is also a Lie groupoid and 
we may define the nonlinear Spencer operator:
\[   \fbox{  $ \bar{D}: {\cal{R}}_{q+1} \rightarrow T^* \otimes R_q : f_{q+1} \rightarrow  f^{-1}_{q+1} \circ j_1(f_q ) - {id}_{q+1} = {\chi}_q    $  }   \]
Indeed, $f^{-1}_{q+1} \circ j_1(f_q) $ is a well defined section of $J_1({\cal{R}}_q)$ over the section $f^{-1}_q \circ f_q = id_q $ of ${\cal{R}}_q$, exactly like $id_{q+1}$, and  
$ id^{-1}_q(V({\cal{R}}_q)) = R_q \subset J_q(T = id^{-1}_q (V({\Pi}_q))$. With a slight abuse of language we may say that $f_{q+1} ({\chi}_q) = j_1 (f_q) - f_{q+1}$ and we recall the 
inductive formula:
\[   \fbox{   $  f^k_r {\chi}^r_{\mu, i} + ... + f^k_{\mu + 1_r} {\chi}^r_{,i } = {\partial}_i f^k_{\mu} - f^k_{\mu + 1_i}    $  }  \]
We obtain in particular the following linear combinations of the Spencer operator:  \\
\[  \left\{ \begin{array}{lcl}
{\chi}^k_{,i} &  =  &  g^k_l {\partial}_i f^l - {\delta}^k_i =  A^k_i - {\delta}^k_i  = g^k_l ( {\partial}_i f^r - f^l_i)  \\
{\chi}^k_{j,i} &  =  &  g^k_l ({\partial}_i f^l_j - A^r_i f^l_{rj}) = g^k_l ({\partial}_i f^l_j - f^l_{ij}) - g^k_l g^r_u f^l_{rj} ({\partial}_i f^u - f^u_i) 
\end{array} \right. \]
and the useful compatibility conditions [31]:  \\
 \[ {\partial}_iA^k_j-{\partial}_jA^k_i-A^r_i{\chi}^k_{r,j}+A^r_j{\chi}^k_{r,i}=0   \]
\[ {\partial}_i{\chi}^k_{l,j}-{\partial}_j{\chi}^k_{l,i}-{\chi}^r_{l,i}{\chi}^k_{r,j}+{\chi}^r_{l,j}{\chi}^k_{r,i}-A^r_i{\chi}^k_{lr,j}+A^r_j{\chi}^k_{lr,i}=0    \]  

\noindent
{\bf PROPOSITION 3.6}: When $det(A)\neq 0$, that is when $\Delta(x)=det({\partial}_if^k(x))\neq 0$ because we have $det(f^k_i(x))\neq 0$ by assumption, there is a {\it nonlinear Spencer sequence} stabilized at order $q$:\\
\[ 0 \longrightarrow {aut(X)} \stackrel{j_q}{\longrightarrow} {\Pi}_q \stackrel{{\bar{D}}_1}{\longrightarrow} C_1(T) \stackrel{{\bar{D}}_2}{\longrightarrow} C_2(T)  \]
where ${\bar{D}}_1$ and ${\bar{D}}_2$ are involutive and a {\it  restricted second nonlinear Spencer sequence}:  \\
\[ 0\longrightarrow \Gamma \stackrel{j_q}{\longrightarrow} {\cal{R}}_q \stackrel{{\bar{D}}_1}{\longrightarrow} C_1 \stackrel{{\bar{D}}_2}{\longrightarrow} C_2   \]
such that ${\bar{D}}_1$ and ${\bar{D}}_2$ are involutive whenever ${\cal{R}}_q$ is involutive. In the case of Lie groups of transformations, the symbol of the involutive system $R_q$ {\it must} be $g_q=0$ providing an isomorphism ${\cal{R}}_{q+1}\simeq {\cal{R}}_q\Rightarrow R_{q+1}\simeq R_q$ and we have therefore $C_r={\wedge}^rT^*\otimes R_q$ for $ r=1,...,n$ like in the linear Spencer sequence.   \\

It remains to graft a variational procedure adapted to the previous results. Contrary to what happens in analytical mechanics or elasticity for example, {\it the main idea is to vary sections but not points}. Hence, we may introduce the variation $\delta f^k(x)={\eta}^k(f(x))$ over the target $y=f(x)$ or set ${\eta}^k(f(x))={\xi}^i(x){\partial}_if^k(x)$ to bring it back over the source $x$ along the "{\it vertical machinery} " but {\it notations like} $\delta x^i={\xi}^i$ {\it or} $\delta y^k={\eta}^k$ {\it have no meaning at all}. \\

As a major result first discovered in specific cases by the brothers Cosserat in 1909 [3] and by Weyl in 1918 [40], we shall prove and apply the following key result:  \\

\noindent
SUCH A PROCEDURE WILL ONLY DEPEND ON THE LINEAR SPENCER OPERATOR AND ITS FORMAL ADJOINT. \\

In order to prove this result, if $f_{q+1},g_{q+1}, h_{q+1} \in {\Pi}_{q+1}$ can be composed in such a way that $f'_{q+1}=g_{q+1}\circ f_{q+1} = f_{q+1} \circ  h_{q+1}$, we get:\\
\[ \begin{array}{rl}
{\bar{D}}f'_{q+1}&=f^{-1}_{q+1}\circ g^{-1}_{q+1}\circ j_1(g_q)\circ j_1(f_q)-id_{q+1}    =   f^{-1}_{q+1}\circ {\bar{D}}g_{q+1}\circ j_1(f_q)+{\bar{D}}f_{q+1} \\
     &= h^{-1}_{q+1}\circ f^{-1}_{q+1}\circ j_1(f_q)\circ j_1(h_q) - id_{q+1}     =    h^{-1}_{q+1} \circ {\bar{D}}f_{q+1} \circ j_1(h_q) + \bar{D} h_{q+1}
 \end{array}    \]
Using the local exactness of the first nonlinear Spencer sequence or ([23], p 219), we may state:  \\
 
\noindent
{\bf LEMMA 3.7}: For any section $f_{q+1}\in {\cal{R}}_{q+1}$, the {\it finite gauge transformation}:\\
\[  \fbox{  $  {\chi}_q \in T^*\otimes R_q  \longrightarrow  {\chi}'_q= f^{-1}_{q+1}\circ {\chi}_q\circ j_1(f_q)+{\bar{D}}f_{q+1} \in T^* \otimes R_q   $  }  \]
exchanges the solutions of the {\it field equations} ${\bar{D}}'{\chi}_q=0$ when $f_{q+1}$ acts on $R_q$ and $j_1(f)$ on $T^*$.  \\

\noindent
We may introduce the {\it formal Lie derivative} on $J_q(T)$ by linearity through the successive formulas:  \\
\[ \{ j_{q+1}(\xi), j_{q+1}(\eta) \} = j_q ([ \xi, \eta ])   \]
\[  L({\xi}_{q+1}){\eta}_q=\{{\xi}_{q+1},{\eta}_{q+1}\}+i(\xi) d{\eta}_{q+1}=[{\xi}_q,{\eta}_q]+i(\eta) d{\xi}_{q+1} \]
\[      (L(j_1({\xi}_{q+1})){\chi}_q)(\zeta)= L({\xi}_{q+1})({\chi}_q(\zeta))-{\chi}_q([\xi,\zeta])   \]

\noindent
{\bf LEMMA 3.8}: Passing to the limit {\it over the source} with ${\chi}_q=\bar{D}f_{q+1}$ and $h_{q+1}=id_{q+1}+t {\xi}_{q+1}+ ... $ for 
$t\rightarrow 0$, we get an  {\it infinitesimal gauge transformation} leading to the {\it infinitesimal variation}:  \\
\[ \fbox{  $    \begin{array}{rcl}
     \delta {\chi}_q & = &  d{\xi}_{q+1}+ L(j_1({\xi}_{q+1})){\chi}_q \\
\delta{\chi}_q(\zeta) & = & i(\zeta) d{\xi}_{q+1} + \{{\xi}_{q+1},{\chi}_{q+1}(\zeta)\} + i(\xi) d{\chi}_{q+1}(\zeta) - {\chi}_q([\xi,\zeta])  \\
                            &  =  & i(\zeta) d{\xi}_{q+1} + L({\xi}_{q+1})({\chi}_q(\zeta)) - {\chi}_q([\xi,\zeta])
\end{array}  $  }    \eqno{(4)}   \]
which {\it only depends on} ${\chi}_q$ but {\it does not depend on} the parametrization of ${\chi}_q$ by $f_{q+1}$. \\

\noindent
{\bf LEMMA 3.9}: Passing to the limit {\it over the target} with ${\chi}_q=\bar{D}f_{q+1}$ and $g_{q+1}=id_{q+1}+ t {\eta}_{q+1}+ ... $ for $t\rightarrow 0$ over the target $Y$, we get the other {\it infinitesimal variation}:\\
\[     \fbox{   $     \delta {\chi}_q= f^{-1}_{q+1}\circ d_Y \, {\eta}_{q+1}\circ j_1(f_q)  $  }     \eqno{(5)}    \]
which {\it highly depends on the parametrization} of ${\chi}_q$. \\

\noindent
{\bf EXAMPLE 3.10}: We obtain for $q=1$:  \\
\[ \begin{array}{ll}
\delta{\chi}^k_{,i}& =({\partial}_i{\xi}^k-{\xi}^k_i)+({\xi}^r{\partial}_r{\chi}^k_{,i}+{\chi}^k_{,r}{\partial}_i{\xi}^r-{\chi}^r_{,i}{\xi}^k_r)   \\
\delta {\chi}^k_{j,i}&=({\partial}_i{\xi}^k_j-{\xi}^k_{ij})+({\xi}^r{\partial}_r{\chi}^k_{j,i}+{\chi}^k_{j,r}{\partial}_i{\xi}^r+{\chi}^k_{r,i}{\xi}^r_j-{\chi}^r_{j,i}{\xi}^k_r-{\chi}^r_{,i}{\xi}^k_{jr})    
 \end{array}  \]
  Introducing the inverse matrix  $B=A^{-1}$, we obtain therefore equivalently:  \\
  \[ \delta A^k_i= {\xi}^r{\partial}_rA^k_i + A^k_r{\partial}_i{\xi}^r - A^r_i{\xi}^k_r \,\,\,  \Leftrightarrow \,\,\,  \delta B^i_k={\xi}^r{\partial}_rB^i_k -B^r_k{\partial}_r{\xi}^i + B^i_r{\xi}^r_k    \]
both with:
\[  \delta {\chi}^k_{j,i}=({\partial}_i{\xi}^k_j - A^r_i{\xi}^k_{jr})+({\xi}^r{\partial}_r{\chi}^k_{j,i}+{\chi}^k_{j,r}{\partial}_i{\xi}^r+{\chi}^k_{r,i}{\xi}^r_j-{\chi}^r_{j,i}{\xi}^k_r)      \]

The explicit general formulas of the three previous lemmas cannot be found somewhere else (The reader may compare them to the ones obtained in [39] by means of the so-called " diagonal " method that cannot be applied to the study of explicit examples). The following unusual difficult proposition generalizes well known variational techniques used in continuum mechanics and will be crucially used for applications like the photoelastic beam experiment in [28]:  \\

\noindent
{\bf PROPOSITION 3.11}: The same variation is obtained when ${\eta}_q = f_{q+1}({\xi}_q+{\chi}_q(\xi))$ in which we set ${\chi}_q = \bar{D}f_{q+1}$, a transformation which only depends on $j_1(f_q)$ and is invertible if and only if $det(A)\neq 0$ or, equivalently, $\Delta \neq 0$. We obtain therefore with ${\bar{\xi}}_{q+1} = {\xi}_{q+1} + {\chi}_{q+1}(\xi)$:  \\
\[  \fbox{  $  i(\zeta){\delta} {\chi}_q \equiv \delta {\chi}_q (\zeta)= i (\zeta) d {\bar{\xi}}_{q+1}  - \{ {\chi}_{q+1}(\zeta), {\bar{\xi}}_{q+1} \} , \,\, \forall \zeta \in T         $   }  \]

\noindent
{\it Proof}: First of all, we get $\bar{\xi}=A(\xi)$ for $q=0$, a transformation which is invertible if and only if $det(A)\neq 0$ and thus $\Delta \neq 0$. In the nonlinear framework, we have to keep in mind that there is no need to vary the object $\omega$ which is given but only the need to vary the section $f_{q+1}$ as we already saw, using ${\eta}_q\in R_q(Y)$ {\it over the target} or ${\xi}_q\in R_q$ {\it over the source}. With ${\eta}_q=f_{q+1}({\xi}_q)$, we obtain for example: \\
\[ \begin{array}{rcccl}
  \delta f^k & = & {\eta}^k & = & f^k_r{\xi}^r \\
   \delta f^k_i & = & {\eta}^k_uf^u_i & = & f^k_r{\xi}^r_i+f^k_{ri}{\xi}^r \\
  \delta f^k_{ij} & = & {\eta}^k_{uv}f^u_if^v_j+{\eta}^k_uf^u_{ij} & = & f^k_r{\xi}^r_{ij} + f^k_{ri}{\xi}^r_j+f^k_{rj}{\xi}^r_i+f^k_{rij}{\xi}^r 
  \end{array}  \]
and so on. Introducing the formal derivatives $d_i$ for $i=1,...,n$, we have:  \\
\[  \delta f^k_{\mu} = d_{\mu}{\eta}^k = {\eta}^k_uf^u_{\mu} + ... = f^k_r{\xi}^r_{\mu} + ... + f^k_{\mu +1_r} {\xi}^r       \]
We shall denote by $\sharp({\eta}_q) = d_{\mu} {\eta}^k \frac{\partial}{\partial y^k_{\mu}}\in V({\cal{R}}_q) $ the corresponding vertical vector field, namely:    \\
\[   \sharp({\eta}_q)= 0\frac{\partial}{\partial x^i}+{\eta}^k(y)\frac{\partial}{\partial y^k}+({\eta}^k_u(y)y^u_i)\frac{\partial}{\partial y^k_i}+({\eta}^k_{uv}(y)y^u_iy^v_j+{\eta}^k_u(y)y^u_{ij} )\frac{\partial}{\partial y^k_{ij}}+ ...  \]
However, the standard prolongation of an infinitesimal change of source coordinates described by the horizontal vector field $\xi$, obtained by replacing all the derivatives of $\xi$ by a section ${\xi}_q \in R_q$ over $\xi \in T$, is the vector field: \\
\[   \flat({\xi}_q)={\xi}^i(x)\frac{\partial}{\partial x^i}+ 0\frac{\partial}{\partial y^k} - (y^k_r{\xi}^r_i(x))\frac{\partial}{\partial y^k_i}-(y^k_r{\xi}^r_{ij}(x)+y^k_{rj}{\xi}^r_i(x)+y^k_{ri}{\xi}^r_j(x)) \frac{\partial}{\partial y^k_{ij}}+ ...                                                             \]
It can be proved that $[\flat({\xi})_q,\flat({\xi}'_q]=\flat([{\xi}_q,{\xi}'_q]), \forall {\xi}_q,{\xi}'_q\in R_q$ {\it over the source}, with a similar property for $\sharp(.)$ {\it over the target} [9, 13]. However, $\flat({\xi}_q)$ {\it is not a vertical vector field and cannot therefore be compared to} $\sharp({\eta}_q)$. The main idea is to consider the vertical vector field 
$T(f_q)(\xi) - \flat({\xi}_q)\in V({\cal{R}}_q)$ whenever $y_q=f_q(x)$. \\
Passing to the limit $t\rightarrow 0$ in the formula $g_q\circ f_q=f_q\circ h_q$, we first get $g\circ f = f\circ h \Rightarrow f(x)+t\eta (f(x)) + ... = f(x + t \xi(x) + ... )$. Using the chain rule for derivatives and substituting jets, we get successively:    \\
\[ \delta f^k(x)={\xi}^r{\partial}_r f^k, \hspace{2mm}  \delta f^k_i={\xi}^r{\partial}_rf^k_i + f^k_r {\xi}^r_i,\hspace{2mm}  \delta f^k_{ij}={\xi}^r{\partial}_rf^k_{ij}+f^k_{rj}{\xi}^r_i + f^k_{ri}{\xi}^r_j + f^k_r{\xi}^r_{ij}           \]
and so on, replacing ${\xi}^rf^k_{\mu + 1_r}$ by ${\xi}^r{\partial}_rf^k_{\mu}$ in ${\eta}_q=f_{q+1}({\xi}_q)$ in order to obtain:  \\
\[   \delta f^k_{\mu} = {\eta}^k_rf^r_{\mu} + ... ={\xi}^i({\partial}_if^k_{\mu}-f^k_{\mu+1_i})+f^k_r{\xi}^r_{\mu}+ ... + f^k_{\mu +1_r}{\xi}^r  \]
where the right member only depends on $j_1(f_q)$ when $\mid\mu\mid=q$. \\
Finally, we may write the symbolic formula $f_{q+1}({\chi}_q)=j_1(f_q)-f_{q+1}=\bar{D} f_{q+1}\in T^*\otimes V({\cal{R}}_q)$ in the explicit form:\\
\[        f^k_r{\chi}^r_{\mu,i} + ... +f^k_{\mu +1_r}{\chi}^r_{,i} = {\partial}_if^k_{\mu}-f^k_{\mu +1_i}   \]
Substituting in the previous formula provides ${\eta}_q=f_{q+1}({\xi}_q+ {\chi}_q(\xi))$ and achieves the proof. \\
Checking directly the proposition is not evident even when $q=0$ as we have indeed:  \\
\[ \begin{array}{lclcl}
\delta {\chi}^s_{,i} & = &  g^s_k (\frac{\partial {\eta}^k}{\partial y^u} - {\eta}^k_u){\partial}_if^u  & = & 
 ({\partial}_i{\bar{\xi}}^s-{\bar{\xi}}^s_i) + ({\chi}^s_{r,i}{\bar{\xi}}^r - {\chi}^r_{,i}{\bar{\xi}}^s_r ) \\
                  &   &   &   =  & {\partial}_i (A^s_r{\xi}^r) + (A^r_t {\chi}^s_{r,i}  - A^r_i  {\chi}^s_{r,t} ) {\xi}^t - A^r_i {\xi}^s_r   \\
                  &  &           & = &  ({\partial}_i {\xi}^s - {\xi}^s_i) + ({\xi}^r {\partial}_r {\chi}^s_{,i} + {\chi}^s_{,r} {\partial}_i {\xi}^r - {\chi}^r_{,i} {\xi}^s_r)
  \end{array}   \]                          
 but cannot be done by hand when $q\geq 1$. \\
\hspace*{12cm}  $ \Box $   \\

\noindent
{\bf REMARK 3.12}: In view of its importance, we provide now another proof of this Proposition. \\
Indeed, we have successively: \\
\[  {\eta}^k = {\xi}^r {\partial}_r f^k, \,\, \, {\eta}^k_u f^u_i = f^k_r {\xi}^r_i + {\xi}^r {\partial}_r f^k_i, \, \, \, 
{\eta}^k_{uv} f^u_i f^v_j + {\eta}^k_u f^u_{ij} =  f^k_r {\xi}^r_{ij} + f^k_{ri}{\xi}^r_j + f^k_{rj}{\xi}^r_i + {\xi}^r {\partial}_r f^k_{ij} , ...   \]
and more generally:  \\
\[  \fbox{  $    \delta f^k_{\mu} = d_{\mu} {\eta}^k = {\eta}^k_u f^u_{\mu} +  ...  = {\xi}^i( {\partial}_i f^k_{\mu} - f^k_{\mu + 1_i}) + f^k_r {\xi}^r_{\mu}  + ... +  f^k_{\mu +1_r} {\xi}^r  $  }  \]
However, we have already obtained the inductive formula allowing to construct ${\chi}_q$, namely:  \\
\[    \fbox{  $      f^k_r {\chi}^r_{\mu,i} + ... + f^k_{\mu + 1_r} {\chi}^r_{,i} = {\partial}_i f^k_{\mu} - f^k_{\mu + 1_i}    $  }  \]
Multiplying each side by ${\xi}^i$, subtracting from the previous formula and substituting, we obtain ${\eta}_q = f_{q+1} ({\bar{\xi}}_q) = j_1(f_q)({\xi}_q)$ with ${\bar{\xi}}_q = {\xi}_q + {\chi}_q (\xi) \Rightarrow {\bar{\xi}} = A(\xi)$, a result not evident at first sight. We invite the reader to recover anew the variation $\delta {\chi}_0$ by using 
$\eta = f_1(\bar{\xi}) = j_1(f)(\xi)$ and: \\
\[ {\eta}_1 = f_2({\bar{\xi}}_1) = j_1(f_1)({\xi}_1) \Rightarrow {\eta}^k_u f^u_i = f^k_s {\bar{\xi}}^s_i + f^k_{si} {\bar{\xi}}^s = f^k_s ({\xi}^s_i + {\chi}^s_{i,r} {\xi}^r) + f^k_{si} (A^s_r {\xi}^r) =  
f^k_r {\xi}^r_i + {\xi}^r {\partial}_r f^k_i  \]  

\noindent
{\bf COROLLARY 3.13}: Combining this proposition with the two previous lemmas, we obtain:  \\
\[  \fbox{  $  d_Y {\eta}_{q+1} = f_{q+1} \circ ( d_X  {\xi}_{q+1} - \{ \bar{D} f_{q+2} (\bullet), {\xi}_{q+1} \} \circ j_1(f)^{-1}             $  } \]
whenever ${\eta}_{q+1} = f_{q+2} ({\xi}_{q+1})$ and the reader may check this formula when $f_{q+2} = j_{q+2} (f)$ in a coherent way with the introduction of this paper. \\

For the Killing system $R_1\subset J_1(T)$ with $g_2=0$, these variations are {\it exactly} the ones that can be found in ([3], (49)+(50), p 124 with a printing mistake corrected on p 128) when replacing a $3\times 3$ skew-symmetric matrix by the corresponding vector. {\it The three last unavoidable Lemmas are thus essential in order to bring back the nonlinear framework of finite elasticity to the linear framework of infinitesimal elasticity that only depends on the linear Spencer operator}.\\
For the conformal Killing system ${\hat{R}}_1\subset J_1(T)$, we obtain:    \\
\[ {\alpha}_i={\chi}^r _{r,i}\Rightarrow \delta {\alpha}_i=({\partial}_i{\xi}^r_r-{\xi}^r_{ri})+({\xi}^r{\partial}_r{\alpha}_i+{\alpha}_r{\partial}_i{\xi}^r - {\chi}^s_{,i}{\xi}^r_{rs}) = ({\partial}_i{\xi}^r_r - A^s_i{\xi}^r_{rs}) + ({\alpha}_r{\partial}_i{\xi}^r + {\xi}^r{\partial}_r {\alpha}_i )     \]
\[ {\varphi}_{ij}={\partial}_i{\alpha}_j - {\partial}_j{\alpha}_i \Rightarrow \delta {\varphi}_{ij}= ({\partial}_j(A^s_i{\xi}^r_{rs}) - {\partial}_i(A^s_j{\xi}^r_{rs})) + ({\varphi}_{rj}{\partial}_i{\xi}^r + {\varphi}_{ir}{\partial}_j{\xi}^r + {\xi}^r {\partial}_r {\varphi}_{ij})  \]
These are {\it exactly} the variations obtained by Weyl ([40], (76), p 289) who was assuming implicitly $A=0$ when setting ${\bar{\xi}}^r_r=0\Leftrightarrow {\xi}^r_r=-{\alpha}_i{\xi}^i$ by introducing a connection. Accordingly, ${\xi}^r_{ri}$ is the variation of the EM potential itself, that is the $\delta A_i$ of engineers used in order to exhibit the Maxwell equations from a variational principle ([40], $\S$ 26) but the introduction of the Spencer operator is new in this framework. If $f_1=id_1$, we have ${\chi}_0=0$ and $\delta {\alpha}_i=({\partial}_i{\xi}^r_r - {\xi}^r_{ri}) +  ({\alpha}_r{\partial}_i{\xi}^r + {\xi}^r{\partial}_r {\alpha}_i )$. \\

Then, using the definition of $a$, namely $det(f^k_i)= e^{na}$, we have:   \\
\[  n \delta a= (1/det(f^k_i))\delta det(f^k_i)= g^i_k\delta f^k_i= {\eta}^s_s = g^i_k({\xi}^r{\partial}_rf^k_i + 
f^k_r{\xi}^r_i)=  n \,{\xi}^r{\partial}_ra + {\xi}^r_r      \]
Using the variation $\delta A^s_i={\xi}^r{\partial}_rA^s_i + A^s_r{\partial}_i{\xi}^r - A^r_i{\xi}^s_r = g^s_l ( \frac{\partial {\eta}^l}{\partial y^k} - {\eta}^l_k){\partial}_i f^k$
we finally get: \\
\[  \begin{array}{rcl}
\delta {\alpha}_i & = &  ({\partial}_i{\xi}^r_r - {\xi}^r_{ri}) + ({\xi}^r{\partial}_r {\alpha}_i + {\alpha}_r{\partial}_i{\xi}^r ) - {\chi}^s_{,i}{\xi}^r_{rs}   \\
      &  =  &  ({\partial}_i{\xi}^r_r - A^s_i{\xi}^r_{rs}) + ({\xi}^r{\partial}_r {\alpha}_i + {\alpha}_r{\partial}_i{\xi}^r)   \\
     & = & [ \frac{\partial {\eta}^s_s}{\partial y^k} - ng^r_l (\frac{\partial {\eta}^l}{\partial y^k} - {\eta}^l_k ) a_r ] {\partial}_if^k - 
                   A^r_i f^k_r {\eta}^s_{sk}   \\  
       & = & [(\frac{\partial {\eta}^s_s}{\partial y^k} - {\eta}^s_{sk}) - n\, b_l(\frac{\partial {\eta}^l}{\partial y^k}- {\eta}^l_k)]\frak{\partial}_i f^k  
\end{array}   \]
The terms ${\partial}_i{\xi}^r_r + ({\xi}^r{\partial}_r {\alpha}_i + {\alpha}_r{\partial}_i{\xi}^r)$ of the variation, including the variation of 
$\alpha={\alpha}_idx^i$ as a $1$-form, are {\it exactly} the ones introduced by Weyl in ([40] formula (76), p 289). We also recognize the variation 
$\delta A_i$ of the $4$-potential used by engineers now expressed by means of second order jets.  \\
The variation over the target is only depending on the components of the Spencer operator, in a coherent way with the general variational formulas that could have been used otherwise. We notice that these formulas, which have been obtained with difficulty for second order jets, could not even be obtained by hand for third order jets. They  show the importance and usefulness of the general formulas providing the Spencer non-linear operators for an arbitrary order, in particular for the study of the conformal group which is defined by second order lie equations with a $2$-acyclic symbol. It is also important to notice that, setting $b(f(x) = a(x), a-i = f^k_i b_k$, we get:\\
 \[{\alpha}_i = n ({\partial}_i a - A^r_i a_r) = n(\frac{\partial b}{\partial y^k} - b_k){\partial}_i f^k={\beta}_k{\partial}_i f^k  \, \Rightarrow \,
 {\varphi}_{ij} =  {\partial}_i {\alpha}_j - {\partial}_j {\alpha}_i = ( \frac{ \partial {\beta}_l}{\partial y^k} - \frac{\partial {\beta}_k}{\partial y^l}) {\partial}_i f^k {\partial}_j f^l      \]
a formula showing that the EM field does not depend on the exchange of source and target.  \\
Using the transformation ${\eta}_1 = f_2 ({\xi}_1)$ with $f_2 \in \hat{\cal{R}}_2$ and ${\xi}_1, {\eta}_1 \in {\hat{R}}_1$ we obtain the contraction 
${\eta}^k = f^k_i {\xi}^i, {\eta}^k_k = {\xi}^r_r + n a_i {\xi}^i$. The inverse transformation allowing to describe ${\hat{R}}^*_1$ is thus 
${\xi}^i = g^i_k {\eta}^k, {\xi}^r_r = {\eta}^k_k - n b_k {\eta}^k$. Using the variational formula $\delta {\chi}_1 = f^{-1}_2 \circ d_Y {{\eta}_2} \circ j_1 (f_1)$, if the dual field of 
$({\chi}^k_{,i}, {\chi}^r_{r,i})$ {\it over the source} is $({\cal{X}}^i_{,k}, {\cal{X}}^i)$, we obtain for the dual field $({\cal{Y}}^{,k}_l, {\cal{Y}}^k)$ {\it over the target}: \\
\[  \Delta \, {\cal{Y}}^{,k}_l = g^s_l {\cal{X}}^{,i}_k {\partial}_i f^k - n b_l {\cal{X}}^i {\partial}_i f^k , \,\,  \Delta {\cal{Y}}^k = {\cal{X}}^i {\partial}_i f^k  \] 
The specific action of the second order jets is thus ${\cal{Y}}^{,k}_l \longrightarrow  {\cal{Y}}^{,k}_l - n b_l {\cal{Y}}^k$ . If we set $f(x)=x, f^k_i = {\delta}^k_i$ and $k=4$, we recognize the transformation $p_i \longrightarrow p_i - e A_i $ because ${\cal{Y}}^4 = {\cal{J}}^4$ is the electric charge density. The factor $a$ can also be used as the additional absolute temperature   
within the framework of the Weyl group and we can also use the full conformal group for introducing the transformation $\Delta {\cal{Y}}^{rs} = {\cal{X}}^{ij} {\partial}_i f^k {\partial}_j f^l $ in order to obtain the extended action of the elations with ${\cal{Y}}^{,k}_l \longrightarrow  {\cal{Y}}^{k}_l - n b_l {\cal{Y}}^k + n \frac{\partial b_l}{\partial y^s} {\cal{Y}}^{ks}$ while taking into account the second set of Maxwell equations $\frac{\partial {\cal{Y}}^{ks} }{\partial y^k} - {\cal{Y}}^s =0 $ in order to obtain the Lorentz force in the right member of the divergence of the generalized Cauchy stress tensor density (See [23] for details).  \\

\noindent
{\bf 4)  CONCLUSION}  \\

Considering a Lie group of transformations as a Lie pseudogroup of transformations, we have revisited in this new framework the mathematical foundations of both continuum mechanics, thermodynamics and electromagnetism, revisiting thus the mathematical foundations of general relativity and gauge theory. As a byproduct, we have proved that the methods known for Lie groups cannot be adapted to Lie pseudogroups and that the two approaches are thus not compatible on the purely mathematical level. \\

In particular, the electromagnetic field, which is a $2$-form with value in the Lie algebra of the unitary group $U(1)$ WHICH IS NOT ACTING ON $X$ according to classical gauge theory, becomes part of a $1$-form with value in a Lie algebroid in the new approach using the Lie pseudogroup of conformal transformations of $X$. More generally, {\it shifting by one step the interpretation of the differential sequences involved},  the "field" is no longer a $2$-form with value in a Lie algebra but must be a $1$-form with value in a Lie algebroid. Meanwhile, we have proved that the use of Lie equations allows to avoid any explicit description of the action of the underlying group, a fact particularly useful for the nonlinear elations of the conformal group. However, a main problem is that the formal methods developed by Spencer and coworkers around 1970 are still not acknowledged by physicists ([33] provides a fine example indeed !) and we don't even speak about the Vessiot structure equations for Lie pseudogroups, not even acknowledged by mathematicians after more than a century [32]. Finally, as a very striking fact with deep roots in homological algebra, the Cauchy/Cosserat/Clausius//Maxwell/Weyl equations can be parametrized, contrary to Einstein equations [27, 36, 37]. We hope this paper will open new trends for future theoretical physics, based on the use of new differential geometric methods (Compare [4] with [9] or [5] with [16]). Accordingly, paraphrasing W. Shakespeare, we may say again as in [19]:  \\

\hspace*{2cm} "TO ACT OR NOT TO ACT, THAT IS THE QUESTION" .  \\  \\

\noindent
{\bf 5) REFERENCES}  \\

\noindent
[1] Bialynicki-Birula, A.: On Galois Theory of Fields with Operators, Am. J. Math., 84 (1962) 89-109.  \\
\noindent
[2] Cauchy, A. L.: Recherches sur l'\'{e}quilibre et le mouvement int\'{e}rieur des corps solides ou fluides, \'{e}lastiques ou non \'{e}lastiques, Bull. Soc. Filomat Paris, 913 (1823) 1,2.  \\
\noindent
[3] Cosserat, E., \& Cosserat, F.: Th\'{e}orie des Corps D\'{e}formables, Hermann, Paris, (1909).\\
\noindent
[4] Janet, M.: Sur les Syst\`{e}mes aux D\'{e}riv\'{e}es Partielles, Journal de Math., 8 (1920) 65-151. \\
\noindent 
[5] Kashiwara, M.: Algebraic Study of Systems of Partial Differential Equations, M\'{e}moires de la Soci\'{e}t\'{e} Math\'{e}matique de France, 63 (1995) (Transl. from Japanese of his $1970$ Master Thesis).  \\
\noindent
[6] Macaulay, F.S.: The Algebraic Theory of Modular Systems, Cambridge Tract 19, Cambridge University Press, London, 1916 (Reprinted by Stechert-Hafner Service Agency, New York, 1964).  \\
\noindent
[7] Maxwell, J.C.: A Treatise on Electricity and Magnetism, Clarendon Press, Oxford, 1873.  \\
\noindent
[8] Poincar\'{e}, H.: Sur une Forme Nouvelle des Equations de la M\'{e}canique, C. R. Acad. Sc. Paris, 132, 7 (1901) 369-371.  \\
\noindent
[9] Pommaret, J.-F.: Systems of Partial Differential Equations and Lie Pseudogroups, Gordon and Breach, New York (1978); Russian translation: MIR, Moscow,(1983).\\
\noindent
[10] Pommaret, J.-F.: La Structure de l'Electromagnetisme et de la Gravitation, C. R. Acad. Sc. Paris, 297 (1983) 493-496.  \\
\noindent
[11] Pommaret, J.-F.: Differential Galois Theory, Gordon and Breach, New York (1983).\\
\noindent
[12] Pommaret, J.-F.: Lie Pseudogroups and Mechanics, Gordon and Breach, New York (1988).\\
\noindent
[13] Pommaret, J.-F.: Partial Differential Equations and Group Theory, Kluwer (1994).\\
http://dx.doi.org/10.1007/978-94-017-2539-2.    \\
\noindent
[14] Pommaret, J.-F.: Dualit\'{e} Diff\'{e}rentielle et Applications. Comptes Rendus Acad. Sciences Paris, S\'{e}rie I, 320 (1995) 1225-1230.  \\
\noindent
[15] Pommaret, J.-F.: Fran\c{c}ois Cosserat and the Secret of the Mathematical Theory of Elasticity, Annales des Ponts et Chauss\'ees, 82 (1997) 59-66 (Translation by D.H. Delphenich).  \\
\noindent
[16] Pommaret, J.-F.: Partial Differential Control Theory, Kluwer, Dordrecht (2001) (Zbl 1079.93001).   \\
\noindent
[17] Pommaret, J.-F.: Algebraic Analysis of Control Systems Defined by Partial Differential Equations, in "Advanced Topics in Control Systems Theory", Springer, Lecture Notes in Control and Information Sciences 311 (2005) Chapter 5, pp. 155-223.\\
\noindent
[18] Pommaret, J.-F.: Parametrization of Cosserat Equations, Acta Mechanica, 215 (2010) 43-55.\\
http://dx.doi.org/10.1007/s00707-010-0292-y.  \\
\noindent
[19] Pommaret, J.-F.: Spencer Operator and Applications: From Continuum Mechanics to Mathematical Physics, in "Continuum Mechanics-Progress in Fundamentals and Engineering Applications", Dr. Yong Gan (Ed.), ISBN: 978-953-51-0447--6, InTech (2012) Available from: \\
http://dx.doi.org/10.5772/35607 .  \\
\noindent
[20] Pommaret, J.-F.: The Mathematical Foundations of General Relativity Revisited, Journal of Modern Physics, 4 (2013) 223-239.
 https://dx.doi.org/10.4236/jmp.2013.48A022.   \\
 \noindent
[21] Pommaret, J.-F.: The Mathematical Foundations of Gauge Theory Revisited, Journal of Modern Physics, 5 (2014) 157-170.  
https://dx.doi.org/10.4236/jmp.2014.55026.  \\
 \noindent
[22] Pommaret, J.-F.: Relative Parametrization of Linear Multidimensional Systems, Multidim. Syst. Sign. Process., 26 (2015) 405-437.  
https://doi.org/10.1007/s11045-013-0265-0.  \\
\noindent
[23] Pommaret, J.-F.: From Thermodynamics to Gauge Theory: The Virial Theorem Revisited, in L. Bailey Editor: " Gauge Theories and Differential Geometry", Nova Science Publishers, New York, 2016. (ISBN 978-1-63483-546-6). Chapter I, p 1-44.  \\
\noindent
[24] Pommaret, J.-F.: Why Gravitational Waves Cannot Exist, Journal of Modern Physics, 8 (2017) 2122-2158.  https://arxiv.org/abs/1708.06575 , 
https://doi.org/104236/jmp.2017.813130.    \\
\noindent
[25] Pommaret, J.-F.: Deformation Theory of Algebraic and Geometric Structures, Lambert Academic Publisher (LAP), Saarbrucken, Germany (2016). A short summary can be found in "Topics in Invariant Theory ", S\'{e}minaire P. Dubreil/M.-P. Malliavin, Springer 
Lecture Notes in Mathematics, 1478 (1990) 244-254. (http://arxiv.org/abs/1207.1964). \\
\noindent
[26] Pommaret, J.-F.: New Mathematical Methods for Physics, Mathematical Physics Books, Nova Science Publishers, New York (2018) 150 pp.  \\
\noindent
[27] Pommaret, J.-F.: Homological Solution of the Lanczos Problems in Arbitrary Dimension, Journal of Modern Physics, 12 
(2021) 829-858.   https://arxiv.org/abs/1803.09610. \\
\noindent
https://doi.org/10.4236/jmp.2020.1110104.  \\
\noindent
[28]  Pommaret, J.-F.: The Mathematical Foundations of Elasticity and Electromagnetism Revisited, Journal of Modern Physics, 10 (2019) 1566-1595.     \\
 https://doi.org/10.4236/jmp.2019.1013104 (https://arxiv.org/abs/1802.02430 ). \\
\noindent
[29] Pommaret, J.-F.: Minimum Parametrization of the Cauchy Stress Operator, Journal of modern Physics, 12 (2021) 453-482. 
https://arxiv.org/abs/2101.03959 \\
\noindent 
https://doi.org/10.4236/jmp.2021.124032.  \\
\noindent
[30] Pommaret, J.-F.: The Conformal Group Revisited, https://arxiv.org/abs/2006.03449.   \\
\noindent
https://doi.org/10.4236/jmp.2021.1213106.  \\
\noindent
[31]  Pommaret, J.-F.: Nonlinear Conformal Electromagnetism, https://arxiv.org/abs/2007.01710. \\
\noindent
https://doi.org/10.4236/jmp.2022.134031.   \\
\noindent
[32] Pommaret, J.-F.: How Many Structure Constants do Exist in Riemannian Geometry ?, Mathematics in Computer Science, (2022) 16:23, 
https://doi.org/10.1007/s11786-022-00546-3.  \\
\noindent
[33] Pommaret, J.-F.: Killing Operator for the Kerr Metric, Journal of Modern Physics, 14 (2023) 31-59. https://arxiv.org/abs/2203.11694 , 
https://doi.org/10.4236/jmp.2023.141003.     \\
\noindent
[34] Pommaret, J.-F.: Gravitational Waves and Parametrizations of Linear Differential Operators,\\ 
https://doi.org/10.5772/intechopen.1000851 (https://doi.org/10.5772/intechopen.1000226).  \\
\noindent
[35] Pommaret, J.-F.: Differential Galois Theory and Hopf Algebras for Lie Pseudogroups, \\
https://arxiv.org/abs/2308.03759.  \\  
\noindent
[36] Pommaret, J.-F.: Control Theory and Parametrizations of Linear Differential Operators, (https://arxiv.org/abs/2311.07779 ).  \\
\noindent
[37] Rotman, J.J.: An Introduction to Homological Algebra, Pure and Applied Mathematics, Academic Press (1979).  \\
\noindent
[38] Spencer, D.C.: Overdetermined Systems of Partial Differential Equations, Bull. Am. Math. Soc., 75 (1965) 1-114.\\
\noindent
[39] Spencer, D.C. and Kumpera, A.: Lie Equations, Princeton University Press, Princeton, 1972.  \\
\noindent
[40] Weyl, H.: Space, Time, Matter, (1918) . New York: Dover; 1952.   \\

\end{document}